\documentclass[a4paper,12pt]{spieman}  %>>> 12pt font mandatory; use this for US letter paper 
\usepackage{amsmath}  %>>> for AMS math formatting, 
\usepackage{graphicx}
\usepackage{color}
\usepackage{amssymb}
\usepackage{setspace}
\usepackage{tocloft}

\title{MuSCAT: a multicolor simultaneous camera for studying atmospheres of transiting exoplanets}

\author{Norio Narita,\supscr{a,b,c} Akihiko Fukui,\supscr{d} Nobuhiko Kusakabe,\supscr{b} Masahiro Onitsuka,\supscr{b,c}\\
Tsuguru Ryu,\supscr{b,c} Kenshi Yanagisawa,\supscr{d} Hideyuki Izumiura,\supscr{c,d} Motohide Tamura,\supscr{a,b,e}\\
Tomoyasu Yamamuro\supscr{f}}

\affiliation{\supscrsm{a}Astrobiology Center, National Institutes of Natural Sciences, 2-21-1 Osawa, Mitaka, Tokyo 181-8588, Japan\\
\supscrsm{b}National Astronomical Observatory of Japan, 2-21-1 Osawa, Mitaka, Tokyo 181-8588, Japan\\
\supscrsm{c}SOKENDAI (The Graduate University of Advanced Studies), 2-21-1 Osawa, Mitaka, Tokyo 181-8588, Japan\\
\supscrsm{d}Okayama Astrophysical Observatory, National Astronomical Observatory of Japan, Kamogata, Asakuchi,
Okayama 719-0232, Japan\\
\supscrsm{e}The University of Tokyo, Department of Astronomy, 7-3-1 Hongo, Bunkyo-ku, Tokyo 113-0033, Japan\\
\supscrsm{f}OptCraft, 3-16-8 Higashi-Hashimoto, Midori-ku, Sagamihara, Kanagawa 252-0144, Japan}

\cftpagenumbersoff{figure}
\cftpagenumbersoff{table} 
\begin{document} 
\maketitle 

%%%%%%%%%%%%%%%%%%%%%%%%%%%%%%%%%%%%%%%%%%%%%%%%%%%%%%%%%%%%% 
\begin{abstract}
We report a development of a multi-color simultaneous camera for the 188cm telescope
at Okayama Astrophysical Observatory in Japan.
The instrument, named MuSCAT, has a capability of 3-color simultaneous imaging in
optical wavelength where CCDs are sensitive.
MuSCAT is equipped with three 1024$\times$1024 pixel CCDs,
which can be controlled independently.
The three CCDs detect lights in $g'_2$ (400--550 nm), $r'_2$ (550--700 nm), and $z_{s,2}$ (820--920 nm)
bands using Astrodon Photometrics Generation 2 Sloan filters.
The field of view of MuSCAT is 6.1$\times$6.1 arcmin$^2$
with the pixel scale of 0.358 arcsec per pixel.
The principal purpose of MuSCAT is to perform high precision multi-color transit photometry.
For the purpose, MuSCAT has a capability of self autoguiding which enables to
fix positions of stellar images within $\sim$1 pix.
We demonstrate relative photometric precisions of
0.101\%, 0.074\%, and 0.076\% in $g'_2$, $r'_2$, and  $z_{s,2}$ bands, respectively,
for GJ436 (magnitudes in $g'$=11.81, $r'$=10.08, and $z'$=8.66) with 30 s exposures. 
The achieved precisions meet our objective, and the instrument is ready for operation.
\end{abstract}

%>>>> Include a list of up to six keywords after the abstract
\keywords{instrumentation, exoplanets, multicolor, photometry, transits}

%>>>> Include contact information for corresponding author
{\noindent \footnotesize{\bf Address all correspondence to}: Norio Narita, Astrobiology Center / National Astronomical Observatory of Japan, \\ 2-21-1 Osawa, Mitaka, Tokyo 181-8588, Japan; Tel: +81 422-34-3543; Fax: +81 422-34-3527;\\
E-mail:  \linkable{norio.narita@nao.ac.jp} }
%%%%%%%%%%%%%%%%%%%%%%%%%%%%%%%%%%%%%%%%%%%%%%%%%%%%%%%%%%%%%

\begin{spacing}{2}   % use double spacing for rest of manuscript

%%%%%%%%%%%%%%%%%%%%%%%%%%%%%%%%%%%%%%%%%%%%%%%%%%%%%%%%%%%%%
\section{Introduction}
\label{sect:intro}  % \label{} allows reference to this section

Transiting planets, which transit in front of their host stars, are especially
important research objects among exoplanets, as when combined with radial velocity measurements,
they can provide us various information about the nature of exoplanets such as
the mass, radius, density, orbital obliquity, and atmosphere.
Most of transiting planets have been discovered by transit surveys which monitor the brightness
of hundreds of thousands of stars.
Several groups have worked or been working on ground-based transit surveys
\cite{2007ApJ...656..552B,2007MNRAS.375..951C,2012ApJ...761..123S},
and CoRoT\cite{2006cosp...36.3749B} and Kepler\cite{2010Sci...327..977B} have performed space-based transit surveys.
Moreover, the second epoch mission of Kepler, namely K2\cite{2014PASP..126..398H}, is now ongoing,
and next generation space missions TESS\cite{2015JATIS...1a4003R} and
PLATO\cite{2014ExA....38..249R} have been approved to be launched around 2017 and 2024, respectively.

The biggest problem for transit surveys, especially for ground-based ones, is
commingling of false positives with candidates of transiting planets.
The cause of false positives of transiting planets are eclipsing binaries, as they mimic
planetary transits by grazing or by being mixed in another bright star's brightness.
Thus it is important to conduct follow-up observations after transit surveys
in order to validate candidates as true planets by eliminating false positives.

High precision multi-color transit photometry is known to be useful for such follow-up observations
to discriminate whether transit-like dimming is caused by a true planet or by an eclipsing binary
\cite{2012MNRAS.426..342C}.
This is because a true planet is almost dark in all wavelength, while an intervening body
of an eclipsing binary is bright itself and its brightness changes significantly with wavelength.
Thus false positives caused by eclipsing binaries can be spotted by observing significant
wavelength dependence in transit depths.
On the other hand, transit depths of a true planet also have wavelength dependence.
Most of the wavelength dependence comes from the stellar limb-darkening,
but the apparent planetary radius also has weak wavelength dependence
which reflects the nature of its planetary atmosphere.
High precision multi-color transit photometry is known to be useful
to measure the weak wavelength dependence in transit depths to study atmospheres of transiting planets.
This kind of study is known as transmission spectroscopy, and numbers of
multi-color transit observations for this purpose have been reported so far
\cite{2013A&A...553A..26N,2013MNRAS.436....2M,2014MNRAS.443.2391M,2014MNRAS.437.1511B,
2013PASJ...65...27N,2013ApJ...773..144N,2013ApJ...770...95F,2014ApJ...790..108F}.

Multi-color simultaneous cameras are very fruitful for the studies described above for two reasons.
First, multi-color simultaneous cameras provide not only higher efficiency but also
more feasibility to accomplish aimed studies than single-color cameras,
since observable transits from a specific ground site are very limited.
Second, simultaneity of multi-color transit photometry is important to avoid systematic
differences of transit depths due to luminosity change in host stars possibly
caused by existence of starspots, plages, stellar activity, and so on. 
For the reasons, multi-color simultaneous cameras such as
GROND\cite{2013A&A...553A..26N,2013MNRAS.436....2M},
BUSCA\cite{2014MNRAS.443.2391M},
ULTRACAM\cite{2014MNRAS.437.1511B},
SIRIUS\cite{2013PASJ...65...27N,2013ApJ...773..144N},
and MITSuME\cite{2013ApJ...770...95F,2014ApJ...790..108F} have been actively used for transit observations.

Considering the fact that more interesting transiting planets will be discovered in the near future by
advanced ground-based surveys, and also space-based surveys, like
K2, TESS, and PLATO, developments of new multi-color simultaneous cameras are highly desired.
We here report a development of such an astronomical instrument named MuSCAT
(Multi-color Simultaneous Camera for studying Atmospheres of Transiting planets),
which is now installed on the 188cm telescope at Okayama Astrophysical Observatory (OAO) in Japan.

The rest of this paper is organized as follows.
We first describe designs of the optical system of MuSCAT and its components (Sec. \ref{sect:design}),
and introduce the control system of MuSCAT (Sec. \ref{sect:control}).
We then report characteristics and performances of MuSCAT shown
in engineering observations (Sec. \ref{sect:test}).
We discuss on some capabilities for future upgrade of MuSCAT (Sec. \ref{sect:discussions}),
and finally summarize this paper (Sec. \ref{sect:summary}).

\section{Optical System}
\label{sect:design}  % \label{} allows reference to this section

\subsection{Scientific Requirements and Design Policies}
\label{subsec:requirements}

We have designed MuSCAT considering the following conditions.
As we plan to use MuSCAT for validations of transiting planets discovered by transit surveys,
at least 2 colors are necessary to discriminate eclipsing binaries from transiting planets.
Considering the cost and available research grants,
we adopt a design for a 3-color simultaneous camera 
with the 3 colors in optical wavelength where CCDs are sensitive.
For transit observations, it is important to obtain good comparison stars in the field of view (FOV)
to achieve high precision transit photometry.
For the reason, we have designed the MuSCAT FOV as wide as possible for the 188cm telescope.
We have also took care of the throughput (TP) of the instrument to achieve high photometric precision.
To achieve higher sensitivity to the utmost extent, we have carefully selected and designed
the MuSCAT optical system including astronomical bandpass filters, dichroic mirrors, and CCDs.

\subsection{Optical Design}
\label{subsec:overview}

%------------- 
   \begin{figure}
   \begin{center}
   \begin{tabular}{c}
   \includegraphics[width=16cm]{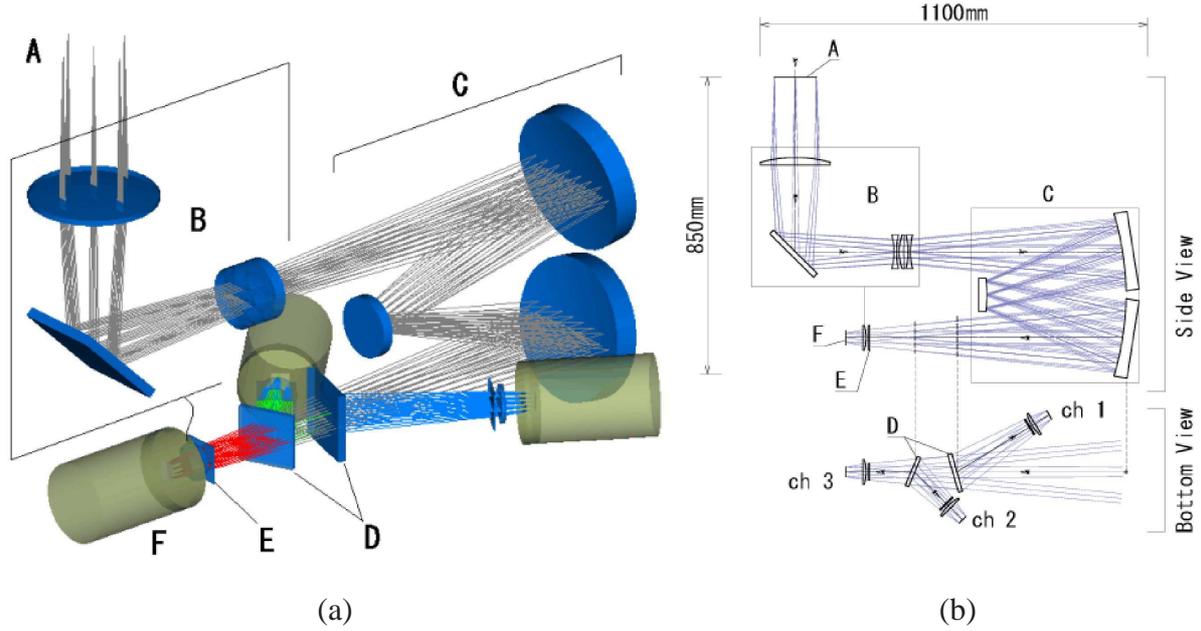}  % fig2 includes two images 
     \\
    \hspace{1cm} (a) \hspace{7.5cm} (b)
   \end{tabular}
   \end{center}
   \caption 
   { \label{layout} %>> use \label inside caption to get Fig. number with \ref{}
Layouts of the optical system of MuSCAT: (a) in 3D and (b) in 2D.
Meanings of the labels in the panels are as follows.
  A : Telescope focal plane (126mm×126mm).
  B : F conversion optics,
      constructed in order with S-BSL7 lens, 45deg mirror,
      CaF2 lens, S-LAL8 lens, CaF2 lens, and
      finally BK7 lens after a bandpass filter in front
      of a detector.
  C : Offner relay.
  D : Dichroic mirrors.
  E : Bandpass filters.
  F : Detectors.
The vertical dashed lines in panel (b) indicate the same positions, but the bottom shows a view from the bottom.
} 
   \end{figure} 
%------------- 

\begin{figure}
   \begin{center}
   \begin{tabular}{c}
   \includegraphics[width=8.5cm]{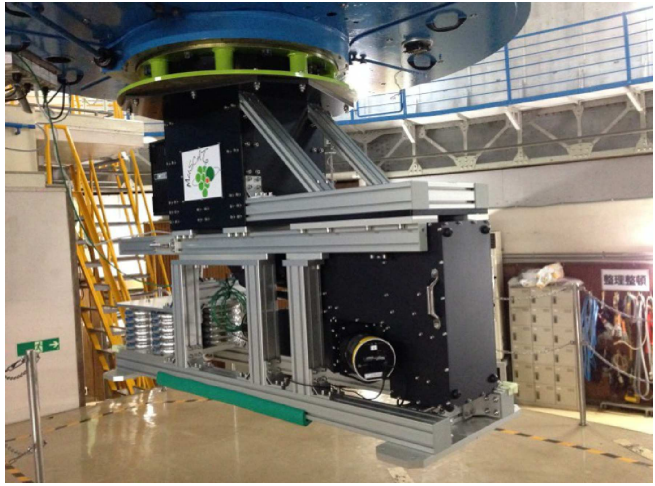}  % fig2 includes two images 
   \end{tabular}
   \end{center}
   \caption 
   { \label{muscatpic} %>> use \label inside caption to get Fig. number with \ref{}
A picture of MuSCAT installed on the 188cm telescope.} 
   \end{figure}

Layouts of the optical system of MuSCAT are shown in Fig. \ref{layout}.
MuSCAT adopts a 45$^{\circ}$ plane mirror and an Offner relay,
which consist of SiO$_2$-protected aluminum mirrors,
and inserts 2 dichroic mirrors in the light path to take images of 3 bands
in optical wavelength simultaneously.
MuSCAT maintains space for another dichroic mirror around the Cassegrain focus
so as to accommodate near-infrared (NIR) channels as a future upgrade capability.
For high precision transit photometry, it is desired to have the widest possible
FOV to get good comparison stars.
The Offner relay enables us to distribute F conversion lenses around the relay.
The lenses contribute to achieve wider FOV by converting F-number from F18 at the telescope
to F5.5 before the relay and further converting to F4.0 just before CCD cameras after the relay.
Aberration correction is performed by 4 lenses before and 1 lens after the Offner relay, and
finally accomplished by a plane-convex lens just before CCD cameras.
Anti-reflection coating for optical wavelength is applied to the lenses.
The lenses are designed to correct aberration of the whole system including
both the 188cm telescope and MuSCAT.
Thereby the optical system of MuSCAT provides a good imaging quality
throughout the FOV.
Wavelength divisions are performed by 2 dichroic mirrors after the Offner relay.
The dichroic mirrors are wedge-shaped to reduce astigmatism.
Astronomical bandpass filters are inserted just before the last lenses.
Lights of astronomical objects are divided into 3 colors and detected by 3 CCD cameras.
A picture of the actual MuSCAT installed on the 188cm telescope is presented
in Fig. \ref{muscatpic}.

%------------- 
%------------- 

\subsection{Dichroic Mirrors and Bandpass Filters}
\label{DM}

\begin{figure}
   \begin{center}
   \begin{tabular}{c}
   \includegraphics[width=10cm]{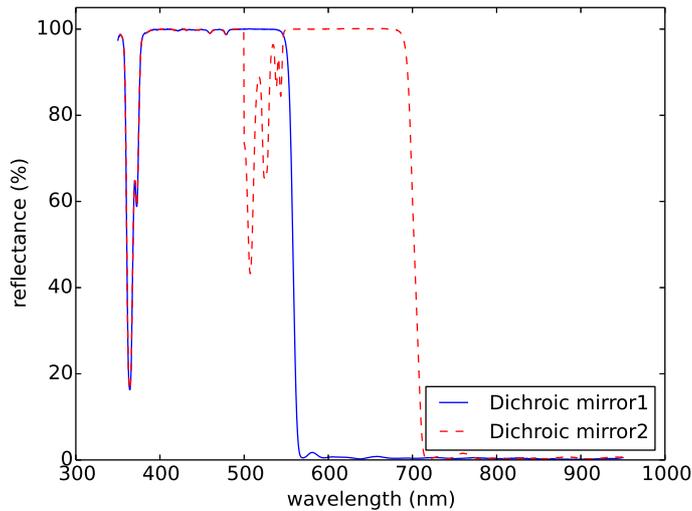}  % fig2 includes two images 
   \end{tabular}
   \end{center}
   \vspace{3cm}
   \caption 
   { \label{DMreflectance} %>> use \label inside caption to get Fig. number with \ref{}
Wavelength dependence of reflectance of DM1 and DM2.} 
   \end{figure}

The 2 dichroic mirrors are manufactured by Asahi Spectra Co.,Ltd.
The size of the first dichroic mirror (DM1) is 113 mm by 108 mm, and the depth
of 11.8 mm with the wedge angle of 7 min 52 sec.
The size of the second dichroic mirror (DM2) is 90 mm by 88 mm,
and the depth of 9.9 mm with the wedge angle of 12 min 33 sec.
Anti-reflection coating is processed on the back sides of the dichroic mirrors so that
the DMs transmit remaining lights (namely, not reflected ones) almost completely.
Fig. \ref{DMreflectance}  plots wavelength dependence of
reflectance of DM1 and DM2 measured by a gonio-spectrophotometer.
The reflectance of both DMs is almost flat across the relevant wavelength for the 3 bands.
We note that there are wiggles in reflectance around 350 nm for both DMs and around 525 nm for the DM2,
but those wiggles do not affect to the current 3 bands.

The DM1 is inserted with an incident angle of 16 deg and 
it reflects most of light in 400-550 nm and transmits remaining light longward of 550 nm.
Reflected light from the DM1 is detected by the ch~1 CCD camera.
Then the DM2 is inserted with an incident angle of 22.5 deg and
it reflects most of light in 550-700 nm and transmits remaining light longward of 700 nm.
Reflected light from the DM2 is detected by the ch~2 CCD camera, and transmitted light
from the DM2 is detected by the ch~3 CCD camera.

\begin{figure}
   \begin{center}
   \begin{tabular}{c}
   \includegraphics[width=10cm]{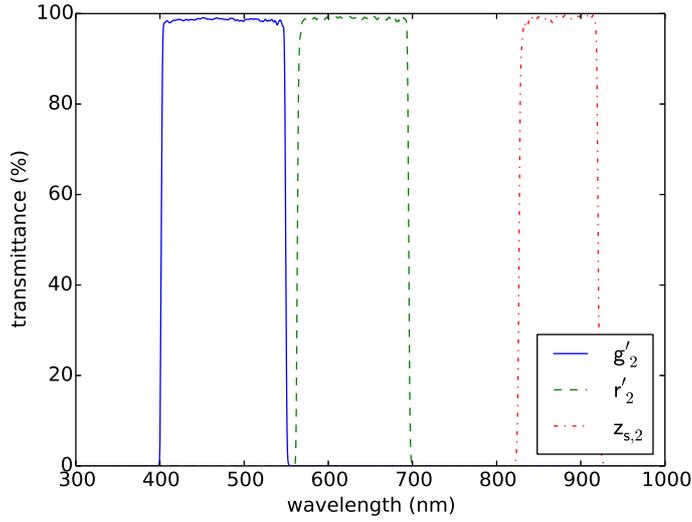}
   \end{tabular}
   \end{center}
   \vspace{3cm}
   \caption 
   { \label{filterproperty} %>> use \label inside caption to get Fig. number with \ref{}
Wavelength dependence of transmittance of bandpass filters.} 
   \end{figure}

We adopt $g'_2$, $r'_2$, $z_{s,2}$ band filters of Astrodon Photometrics Generation 2 Sloan filters.
The size of those filters are 50 mm by 50 mm.
Fig. \ref{filterproperty} plots the wavelength dependence of transmittance
of the bandpass filters, measured by a spectrophotometer SHIMADZU UV-3100PC.
The reflectance and transmittance of DM1 and DM2 are optimized for those 3 bands.

We note that changes in the reflectance and transmittance of DMs and bandpass filters are negligible
in normal operations unless dew condensations occur.

%------------- 
%------------- 

%------------- 
%------------- 

%------------- 
   \begin{figure}
   \begin{center}
   \begin{tabular}{c}
   \includegraphics[width=10cm]{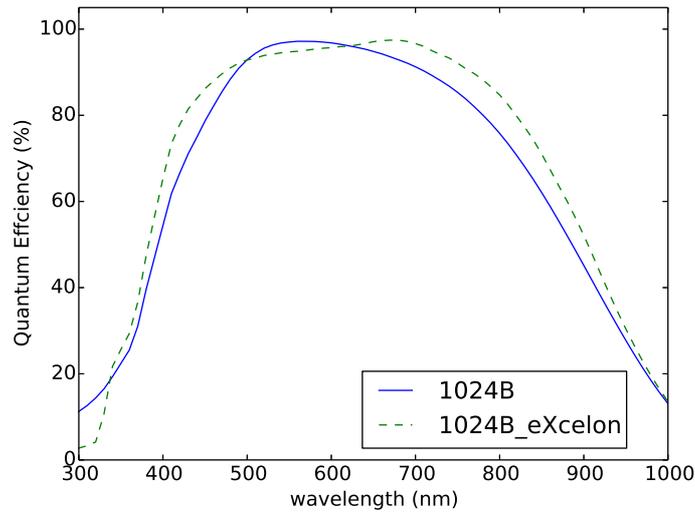}
   \end{tabular}
   \end{center}
   \vspace{3cm}
   \caption 
   { \label{ccdefficiency} %>> use \label inside caption to get Fig. number with \ref{}
Wavelength dependence of quantum efficiency of CCDs.} 
   \end{figure} 
%------------- 

%------------- 
   \begin{figure}
   \begin{center}
   \begin{tabular}{c}
   \includegraphics[width=10cm]{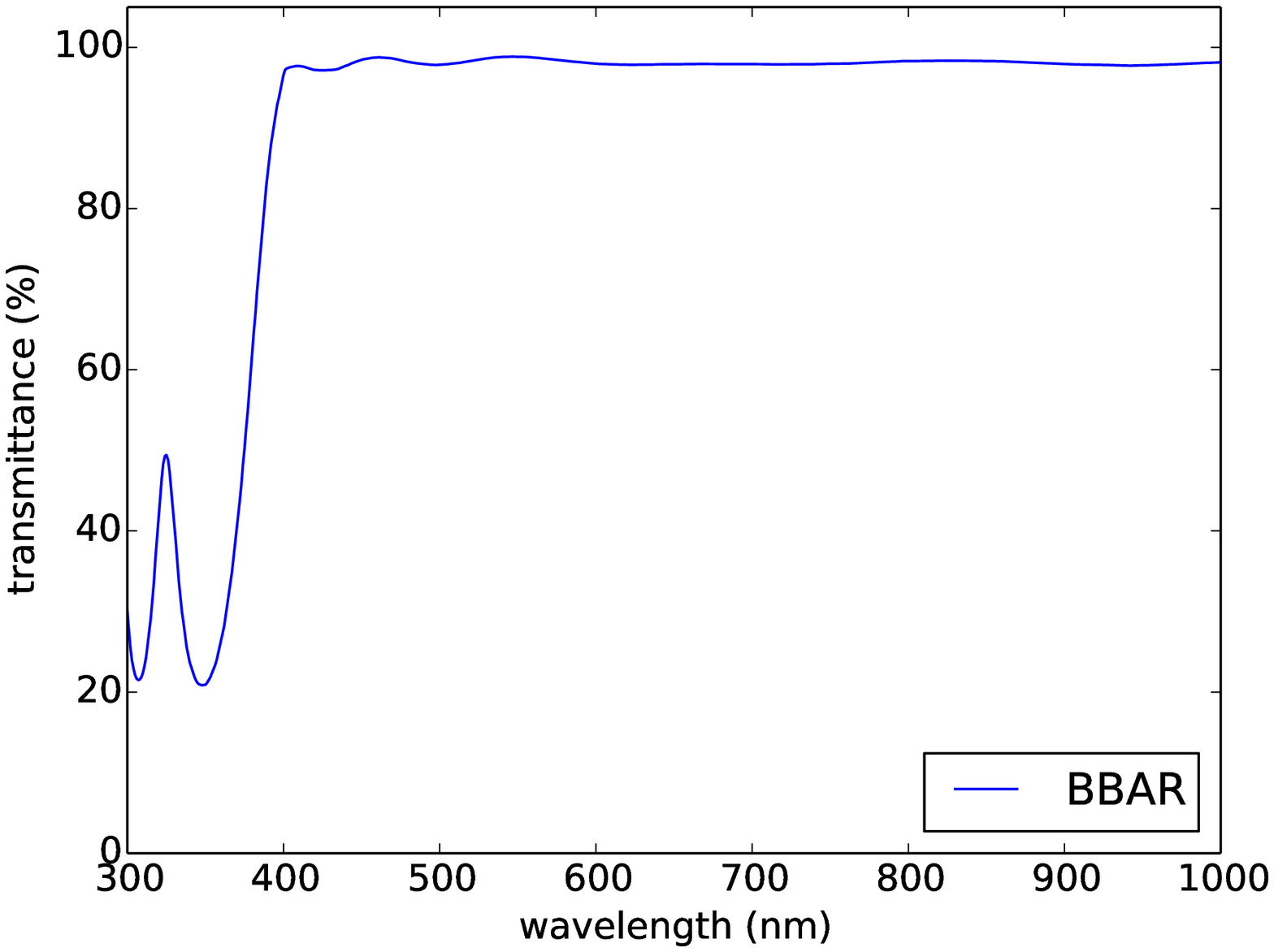}
   \end{tabular}
   \end{center}
   \vspace{2.8cm}
   \caption 
   { \label{ccdcoating} %>> use \label inside caption to get Fig. number with \ref{}
Wavelength dependence of transmittance of BBAR coating.} 
   \end{figure} 
%------------- 

\subsection{Cameras and Detectors}

%% Table 1

MuSCAT equips 3 CCD cameras manufactured by Princeton Instruments.
The first one is a PIXIS:~1024B model, used as the ch 2 CCD camera in $r'_2$ band.
The other two are PIXIS:~1024B\_eXcelon model cameras, used as the ch~1 ($g'_2$ band)
and ch~3 ($z_{s,2}$ band) CCD cameras.
Each CCD camera equips back-illuminated grade 1 CCD chip
(e2v CCD47-10 for PIXIS:~1024 and Princeton Instruments' proprietary CCD for PIXIS:~1024B\_eXcelon)
with 1k$\times$1k (1024$\times$1024) pixels.
It is noted that the CCDs of PIXIS:~1024B\_eXcelon are specially-processed to
suppress the etaloning (fringing) that occurs in standard back-illuminated CCDs.
Quantum efficiencies (QE) of PIXIS:~1024B and PIXIS:~1024B\_eXcelon are plotted in Fig. \ref{ccdefficiency}.
Special anti-reflection coating (BBAR coating) is applied to the vacuum windows of CCD cameras
and its transmittance is shown in Fig. \ref{ccdcoating}.
The data of the QEs and the transmittance of the BBAR coating are provided by the manufacturer.
We summarize nominal specifications of the CCD cameras in Table \ref{CCDsummary}.
We also present actual measured values of gains, full well, and read noise as well as an upper limit of
dark current based on data taken during engineering observations (Sec. \ref{sect:test})
in Table \ref{CCDsummary}.

%------------- 
   \begin{figure}
   \begin{center}
   \begin{tabular}{c}
   \includegraphics[width=10cm]{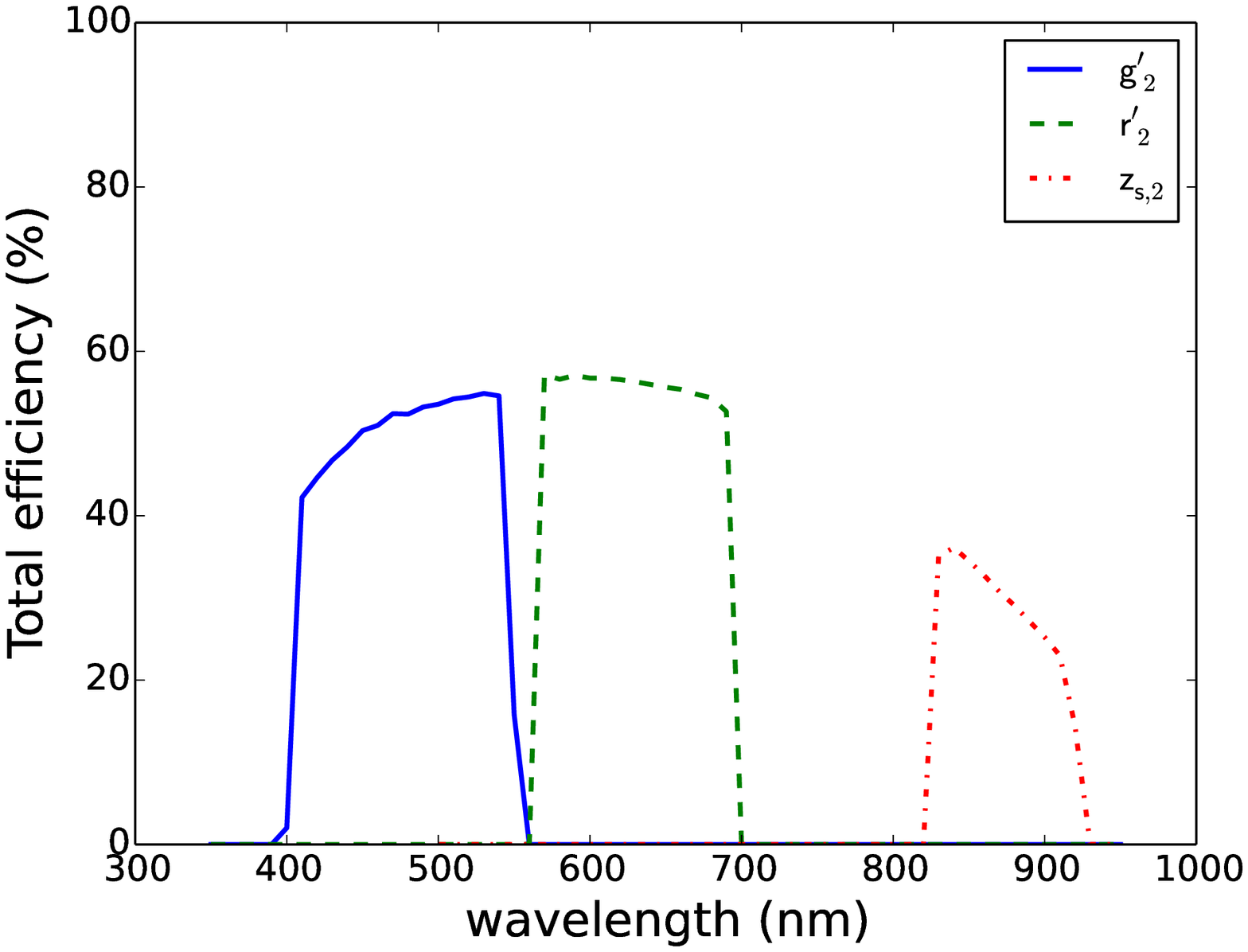}
   \end{tabular}
   \end{center}
   \vspace{2.8cm}
   \caption 
   { \label{totalefficiencyfig} %>> use \label inside caption to get Fig. number with \ref{}
Wavelength dependence of expected total efficiencies of MuSCAT in $g'_2$, $r'_2$, and $z_{s,2}$ bands.} 
   \end{figure} 
%------------- 

\subsection{Total Throughput}

Efficiencies of the F conversion optics and the Offner relay are
roughly estimated as 60\% in $g'_2$ band, 61\% in $r'_2$ band, and 50\% in $z_{s,2}$ band.
Based on the transmittance and reflectance of DMs, filters, BBAR coating, and QE of CCDs
in the previous subsections, we calculate expected total throughput of MuSCAT.
We plot the wavelength dependence of the expected total throughput of MuSCAT
in Fig.~\ref{totalefficiencyfig} and present machine-readable values in Table~\ref{totalefficiencytable}.
A comparison of the expected total throughput with a measured one is presented in section 4.3.

%% Table 2

\subsection{Field of View}

MuSCAT has 6.1$\times$6.1 arcmin$^2$ FOV with the 1k$\times$1k CCDs introduced above.
The pixel scale for each CCD is $\sim$0.358 arcsec per pixel (see Fig~\ref{distm67} for details).
Centers of FOV of 3 CCD cameras are adjusted within 10 pixels in both x and y directions,
and relative rotations of the position angle fit within 1 deg.

\begin{figure}
   \begin{center}
   \begin{tabular}{c}
   \includegraphics[width=12cm]{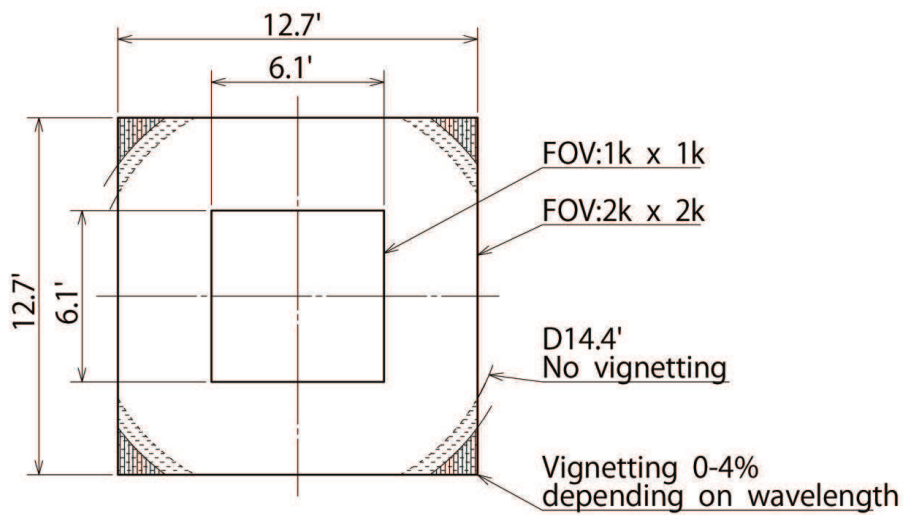}
   \end{tabular}
   \end{center}
   \vspace{0cm}
   \caption 
   { \label{FOV} %>> use \label inside caption to get Fig. number with \ref{}
A schematic view of the potential field of view of MuSCAT.} 
   \end{figure}

As a future upgrade capability, MuSCAT can replace CCD cameras with
2k$\times$2k (2048$\times$2048 pixels) CCD cameras PIXIS:~2048B and PIXIS:~2048B\_eXcelon.
With such an upgrade, the FOV of MuSCAT will get larger to 12.7$\times$12.7 arcmin$^2$.
In that case small (0-4\%) vignetting depending on wavelength would occur in the region
outside the diameter of 14.4 arcmin from center of FOV, while there is no vignetting
for 1k$\times$1k CCDs.
FIg. \ref{FOV} indicates the explained potential FOV of MuSCAT. 

%------------- 
%------------- 

\subsection{Spot Diagram}

\begin{figure}
   \begin{center}
   \begin{tabular}{c}
   \includegraphics[width=16cm]{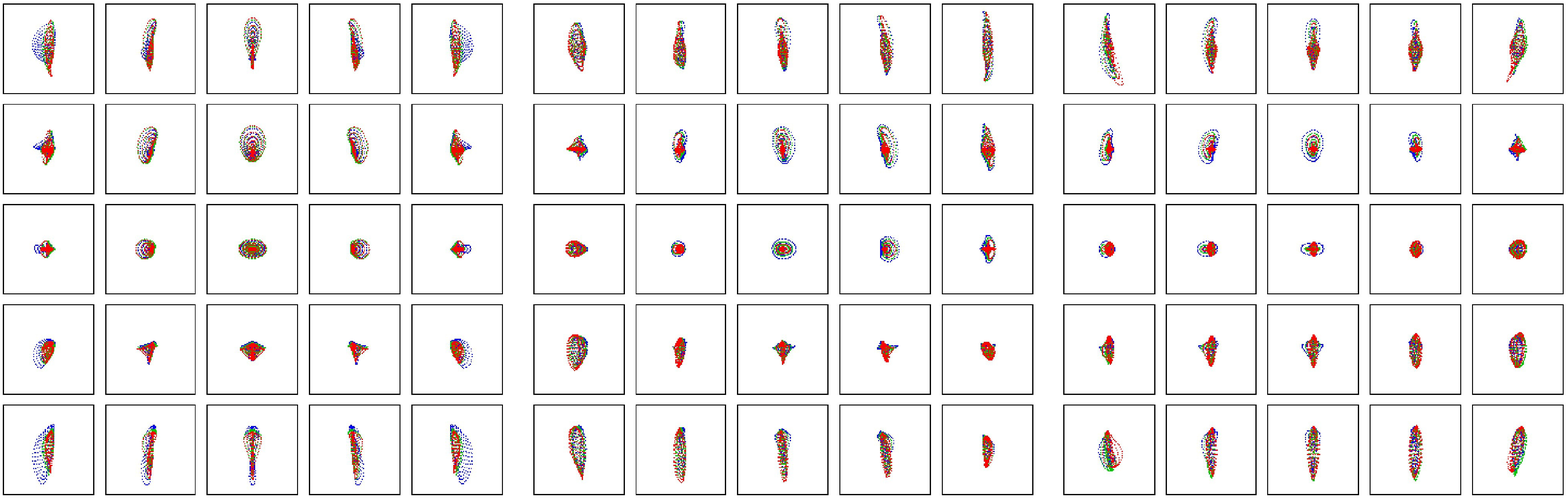}  % fig2 includes two images 
     \\
     (a) \hspace{4.72cm} (b) \hspace{4.72cm} (c)
   \end{tabular}
   \end{center}
   \caption 
   { \label{spot1} %>> use \label inside caption to get Fig. number with \ref{}
Spot diagram of the CCD cameras for on-focus cases for (a) ch~1, (b) ch~2, and (c) ch~3.
The 5$\times$5 cells represent the FOV of 2k$\times$2k CCD case,
while the inner 3$\times$3 cells do a case for 1k$\times$1k CCD.
The size of each cell corresponds to 2$\times$2 arcsec.
Colors indicate simulated images of the shortest (blue), mid (green), and longest (red)
wavelength in each channel.
Specifically, 400, 470, 550 nm for ch~1, 
550, 630, 700 nm for ch~2, and
700, 800, 950 nm for ch~3.
} 
   \end{figure}

\begin{figure}
   \begin{center}
   \begin{tabular}{c}
   \includegraphics[width=16cm]{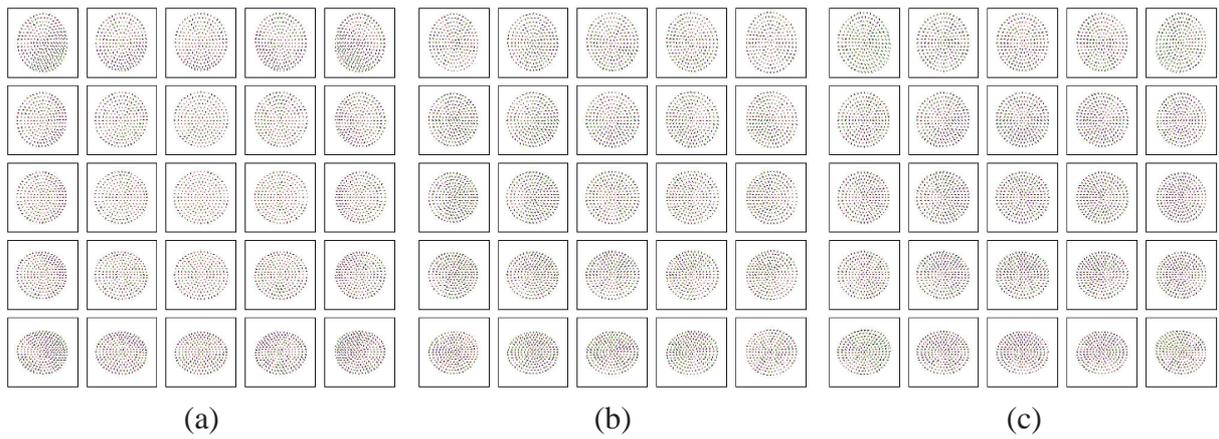}  % fig2 includes two images 
     \\
     (a) \hspace{4.72cm} (b) \hspace{4.72cm} (c)   \end{tabular}
   \end{center}
   \caption 
   { \label{spot2} %>> use \label inside caption to get Fig. number with \ref{}
Same as Fig. \ref{spot1}, but for defocused cases.
The size of each cell corresponds to 10$\times$10 arcsec.} 
   \end{figure}

Simulated imaging performances of the MuSCAT optical system are shown in
spot diagrams in Figs. \ref{spot1} and \ref{spot2}.
Fig. \ref{spot1} plots spot diagrams for on-focus cases which indicate
spot radius of all wavelength are well
less than 1 arcsec throughout the current FOV and the potential FOV.
This imaging performance is thus sufficient for the Okayama Astrophysical Observatory
where the typical seeing is about 1.5 arcsec.
Fig. \ref{spot2} shows spot diagrams for defocused cases where
the secondary mirror is shifted by 1.5 mm, which makes spot radius expand to about 4 arcsec.
It is well known that defocusing is very useful for high precision transit photometry
for isolated sources\cite{2009MNRAS.396.1023S},
and thus those cases are more realistic for transit observations.
The panels imply that images are almost circular throughout the FOV and suitable
for aperture photometry.

%------------- 
%------------- 

%------------- 
%------------- 

\section{Control System}
\label{sect:control}  % \label{} allows reference to this section

\subsection{System Structure}

\begin{figure}
   \begin{center}
   \begin{tabular}{c}
   \includegraphics[width=12cm]{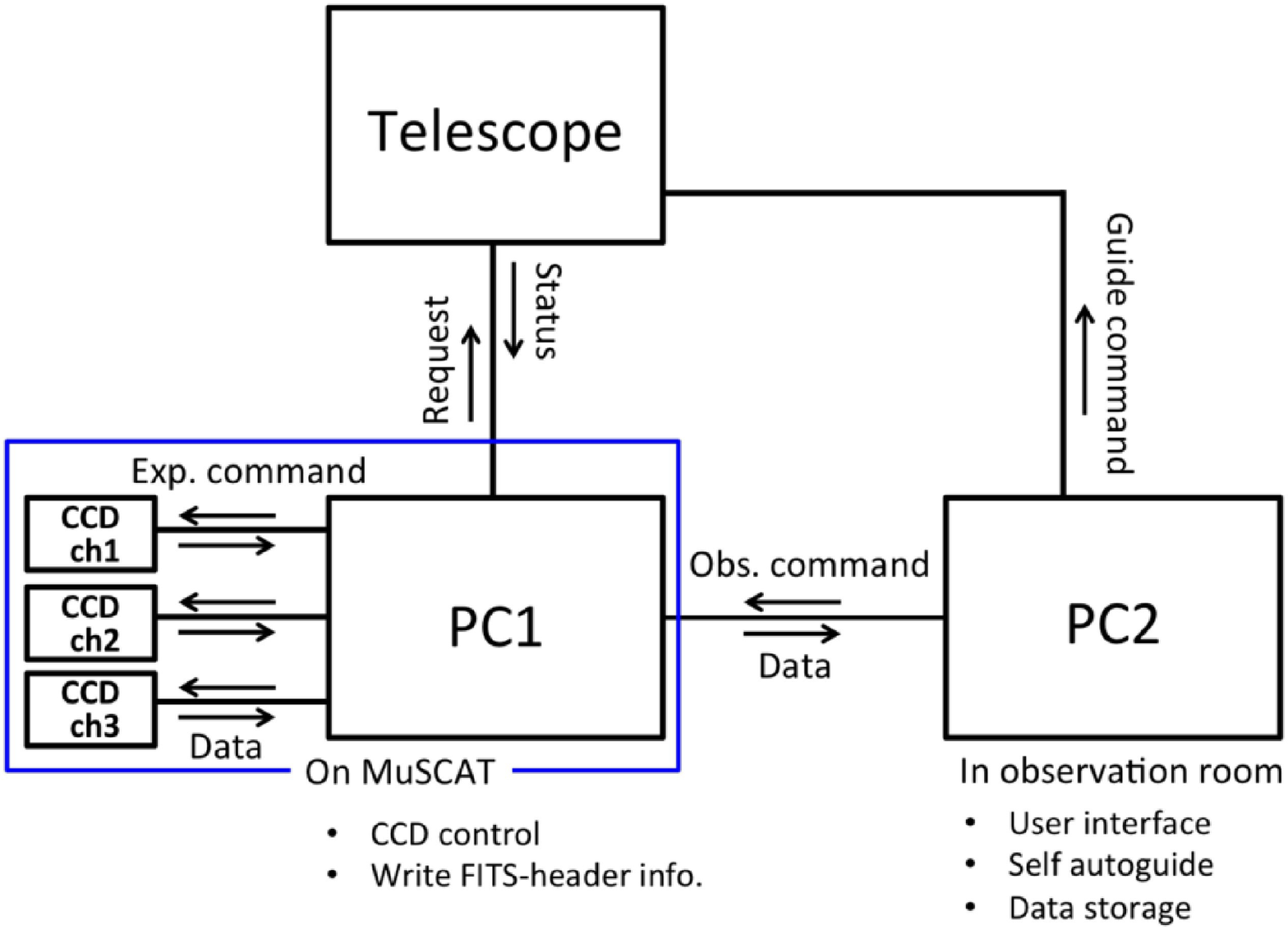} 
   \end{tabular}
   \end{center}
   \caption 
   { \label{control} %>> use \label inside caption to get Fig. number with \ref{}
A flow chart of MuSCAT control system.} 
   \end{figure}

Fig. \ref{control} shows a flow chart of the MuSCAT control system.
The system consists of 2 personal computers: one (PC1) is directly mounted on the instrument
and the other (PC2) is located in an observation room in the observatory.
We use Linux/CentOS 5 (32bit) for PC1 and Linux/CentOS 7 (64bit) for PC2.
For PC1, a 32bit operating system is required to customize observation commands
using 32bit library for the CCD cameras.
Those operating systems may be subject to change in the future.

Each of the CCD cameras is connected to PC1 with a USB cable.
Each camera is independently controlled by PC1 using an interface software
which is developed based on PVCAM and CFITSIO C libraries.
The images taken by these cameras are saved in a 16-bit FITS format
along with FITS header information including the time information,
telescope status, dome status, and weather information.
For the time information, PC1 refers to the internal clock which is
synchronized to Network Time Protocol servers via the internet.
The typical time offset between the local machine and the NTP servers is less than 1 milliseconds.
Thus total systematic uncertainties on the times of exposures,
including the time lag between the internal time acquisition and actual exposure,
are well within 1 second, which is negligible for even time-critical sciences
such as transit timing measurements.
For the other information than the time, PC1 refers to a telescope-control PC,
which gathers all the up-to-the-second information.

PC2 provides a user interface.
Observers can send an observing command to each camera from this PC
specifying the exposure time, number of images,
gain and readout speed settings, and so on.
We note that one can select optimal settings for each camera.
Namely, one can set a different exposure time and a different readout speed to each camera.
This is an important feature of MuSCAT that we can use
an optimal exposure time for each band independently.
The observed images are instantaneously sent to PC2 to be displayed in SAOImage DS9.
PC2 also has a function of self autoguiding (see Section \ref{subsec:self}),
and has a data storage.

%------------- 
%------------- 

\subsection{Self Autoguiding}
\label{subsec:self}

To achieve a 0.1\% level high precision relative photometry,
it is essential to receive stellar fluxes by the same pixels
during a set of observations to mitigate the incompleteness of
the pixel-to-pixel sensitivity correction.
Because the tracking of the 188cm telescope is not perfect,
an autoguiding system is critical to keep the stars on the same positions on the detectors.
However, MuSCAT has no guide camera to capture a guide star located
in the surrounding area of the FOV of the science cameras.
We therefore have developed a self autoguiding system which uses scientific images for guiding
\cite{2013PASJ...65...27N,2013ApJ...773..144N,2013ApJ...770...95F,2014ApJ...790..108F}.
Specifically, the stellar centroids of several bright stars are measured on
one of the three band images soon after the image is obtained.
Then the mean stellar displacement on the latest image relative to
a reference image is calculated to feed back to the telescope.
All the above processes are done within a few seconds after the end of exposure.
We note that the CCD channel to be used for the stellar centroid calculation is selectable.
We usually set the exposure time for the guiding channel to 30--60 seconds
such that the feedback result will be well reflected in the next image
and that the guiding frequency is high enough.
As shown in Section~\ref{subsec:precision}, the autoguiding system can stabilize
the stellar centroid positions within $\sim$1 pixel for bright stars (magnitude less than about 12).
This is important to reduce systematic errors caused by incompleteness of flat-fielding.
We find no large difference in the autoguiding performance
between in-focus and out-of-focus observations, meaning that the guiding performance is not limited
by the degree of defocus but limited by the change of stellar PSF shapes due to seeing variation
and the mechanical accuracy of telescope driving (tracking and fine moving).

\section{Results of Engineering Observations}
\label{sect:test}  % \label{} allows reference to this section

We conducted the first light engineering observation on the night of December 24, 2014,
and further engineering observations on the nights of March 2-4 and April 3-5, 2015.
We examine the performance of MuSCAT using data taken on the nights.
We summarize results of the engineering observations below.

\subsection{Detector Characteristics including Bias, Flat, and Linearity}

\begin{figure}
   \begin{center}
   \begin{tabular}{c}
   \includegraphics[width=15cm]{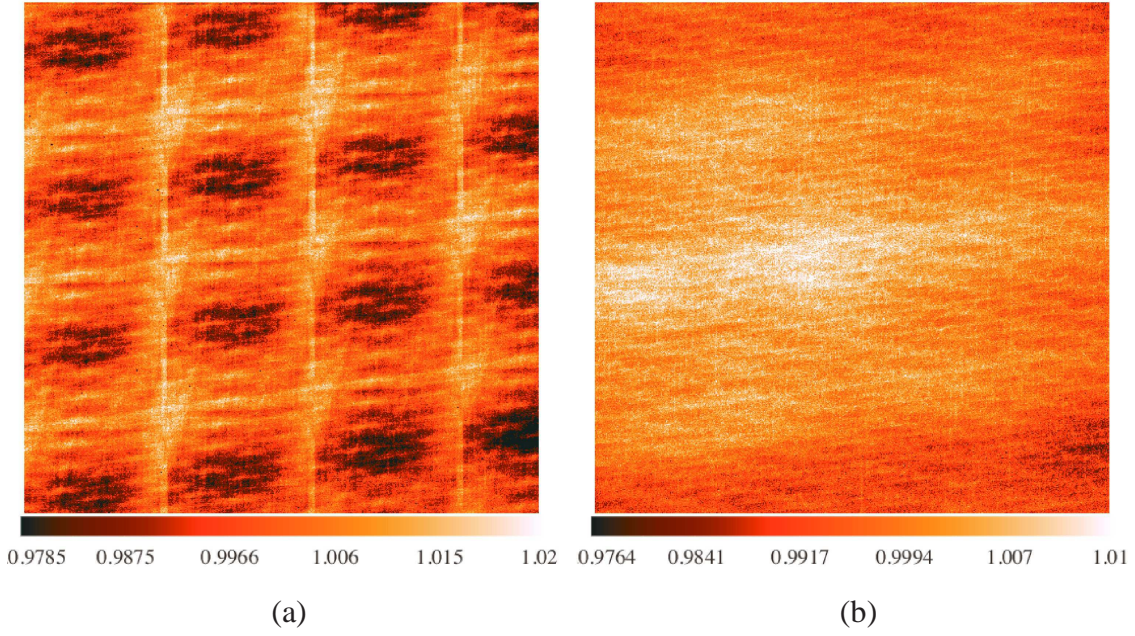} 
    \\
\hspace{-0.2cm} (a) \hspace{6.8cm} (b)
   \end{tabular}
   \end{center}
   \caption 
   { \label{flatgr} %>> use \label inside caption to get Fig. number with \ref{}
Median flat images of (a) $g'_2$ band (with low etaloning process) and
(b) $r'_2$ band (standard CCD).} 
   \end{figure}

We have took hundreds of bias, flat, and linearity test frames during engineering observations
in order to learn detector characteristics of MuSCAT.
We have observed lights of a filament lamp projected onto a matte whiteboard on a wall of
the 188cm telescope dome for flat frames and linearity test frames.
We do not find any strange features or significant bad pixel regions in bias and flat frames,
except a unique feature of low etaloning process in $g'_2$ and $z_{s,2}$ band flat images.
For reference, we present median flat images in $g'_2$ (median of 433 flat frames)
and $r'_2$ (median of 490 flat frames) bands in Fig. \ref{flatgr}.
We note that the unique feature has a very good repeatability with the fractional fluctuation of much less than 0.1\%,
which has little impact on the photometric precision required for our purposes.
We have also exposed a very bright star of $V=6$ to check for the existence of image persistence on the CCDs
and ghost patterns due to reflection by the lenses. We found no apparent features that can affect the photometry.
We have derived full well values, gains, readout noises for the data.
The values are presented in Table \ref{CCDsummary}.

%------------- 
%-------------

%------------- 
   \begin{figure}
   \begin{center}
   \begin{tabular}{c}
   \includegraphics[width=8cm]{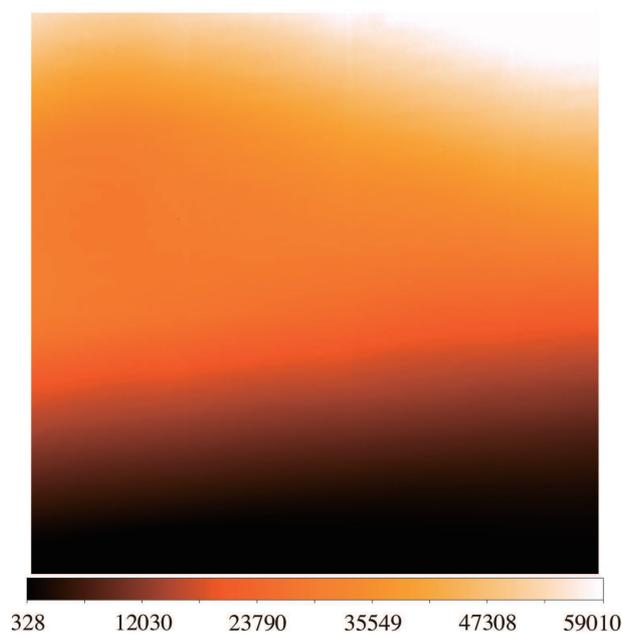} 
   \end{tabular}
   \end{center}
   \caption 
   { \label{linearityframe} %>> use \label inside caption to get Fig. number with \ref{}
An example of a linearity test frame in $g'_2$ band. Counts of the dark region are nearly bias level and 
those in the brightest region are saturated.} 
   \end{figure} 
%-------------

%------------- 
   \begin{figure}
   \begin{center}
   \begin{tabular}{c}
   \includegraphics[width=12cm]{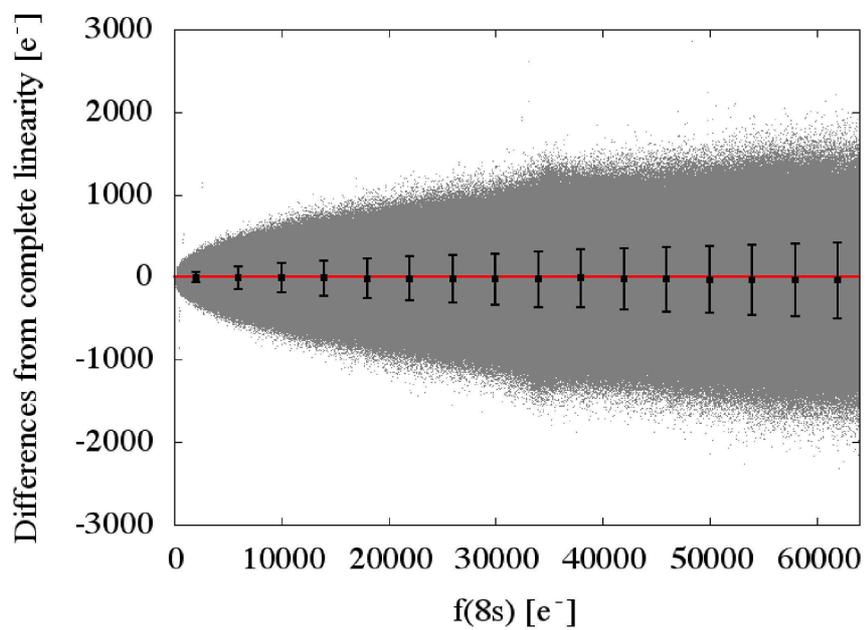} 
   \end{tabular}
   \end{center}
   \vspace{-0.5cm}
   \caption 
   { \label{linearityresult} %>> use \label inside caption to get Fig. number with \ref{}
An example of a linearity test result. The horizontal axis (f(8s) [e-]) indicates electron counts in 8 seconds.
The vertical axis shows the flux ratio parameter which is defined as f(8s) - 2 * f(4s) [e-].
}
   \end{figure} 
%-------------

We have also tested the linearity of MuSCAT CCDs for each readout speed and each gain.
Our method is based on a previous study for the CCDs of High Dispersion Spectrograph (HDS)
of the Subaru telescope\cite{Tajitsu2010}.
First, we have created linearity test frames which have gradational counts on the CCDs,
by opening only a half of the tertiary mirror cover and inserting a black plate into
the light path in front of MuSCAT.
Fig. \ref{linearityframe} shows an example of a linearity test frame.
Second, we monitor counts of the filament lamp until the filament lamp is stabilized.
We note that it takes about 2 hours until counts become nearly-unchanged.
We then start linearity test exposures as follows.
We first determine an exposure time for each CCD which
gives counts from bias level to saturation level gradationally on the CCDs.
We define frames with the above exposure time as ``A'' frames and
ones with a half of the exposure time as ``B'' frames.
We then take A and B frames alternately until obtaining 20 frames each.
We have repeated such exposures for each gain and each readout speed,
namely for the gain modes of 1, 2, 4 e$^-$/ADU,
and for the readout speed of 100kHz and 2MHz.
Subsequently we subtract a median bias frame for each gain and each readout speed.
We then make a new frame which computes photon counts of each pixel in an A frame minus
twice of photon counts for the same pixel in a B frame using adjacent A and B frames
(39 pairs in total for each gain and each readout speed).
We define those frames as ``C'' frames (namely, C = A $-$ 2 $\times$ B for each pixel).
To visualize the linearity of the CCDs,
we plot electron counts (namely, photon counts $\times$ gain) of pixels in A frames as X-axis
and electron counts of the same pixels in C frames as Y-axis.
An example of such a figure is shown in Fig. \ref{linearityresult}.
We finally fit the plotted data with a linear function (Y = $a$X) using the data up to X=64000,
and the best fit linearity slopes are summarized in Table \ref{linearity}.
Based on the above test, we have confirmed that MuSCAT CCDs have a good linearity
within $\sim$0.21\% at a maximum up to the saturation level.
The result means that the effect of non-linearity is well negligible even for
high precision transit photometry if counts of stars do not change drastically
during observations.
In the case we need to correct non-linearity, we will use those data for non-linearity corrections.

%% Table 3

\subsection{PSF and Distortion}

%------------- 
   \begin{figure}
   \begin{center}
   \begin{tabular}{c}
\hspace{-5mm}   \includegraphics[width=17cm]{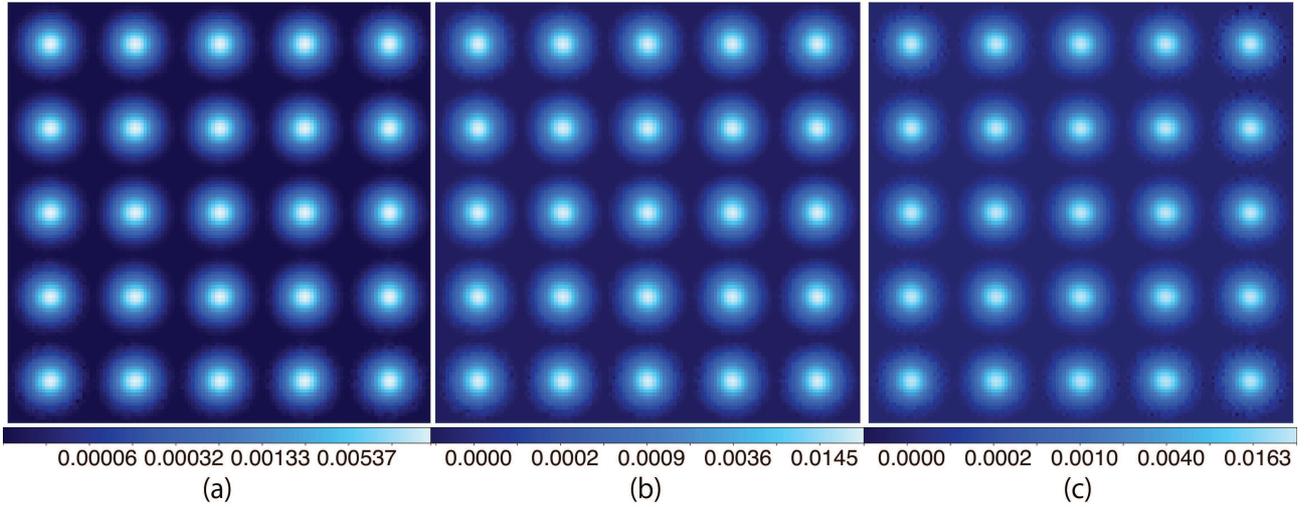} 
%   \\
%\hspace{-5mm} (a) \hspace{5cm} (b) \hspace{5cm} (c)
   \end{tabular}
   \end{center}
   \caption 
   { \label{psfm67} %>> use \label inside caption to get Fig. number with \ref{}
PSF on the detectors extracted from M67 images in (a) $g'_2$ band, (b) $r'_2$ band, and (c) $z_{s,2}$ band.
} 
   \end{figure} 
%-------------

%------------- 
   \begin{figure}
   \begin{center}
   \begin{tabular}{c}
\hspace{-5mm}   \includegraphics[width=17cm]{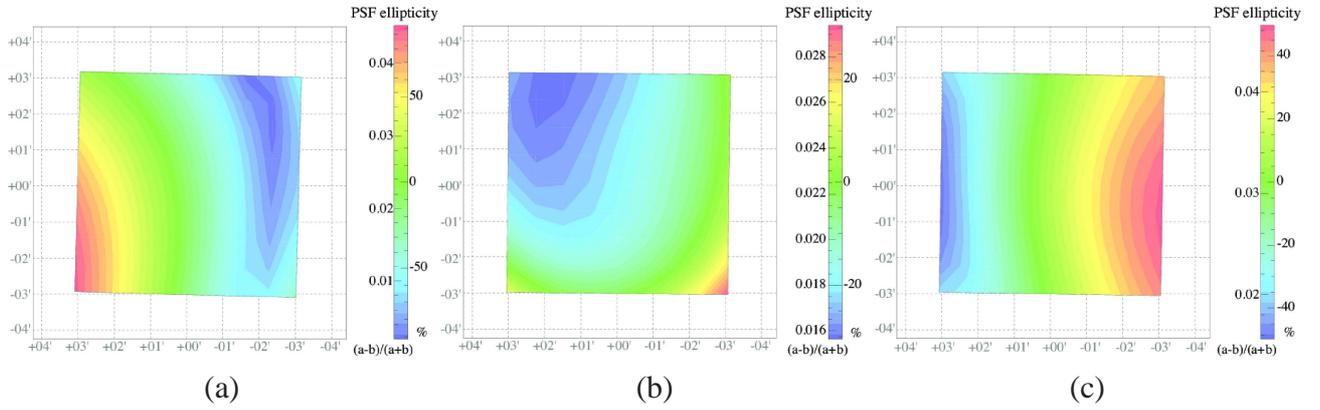} 
    \\
\hspace{-5mm} (a) \hspace{5cm} (b) \hspace{5cm} (c)
   \end{tabular}
   \end{center}
   \caption 
   { \label{ellipm67} %>> use \label inside caption to get Fig. number with \ref{}
Same as Fig. \ref{psfm67}, but ellipticity maps on the detectors.
} 
   \end{figure} 
%-------------

%------------- 
   \begin{figure}
   \begin{center}
   \begin{tabular}{c}
\hspace{-5mm}   \includegraphics[width=17cm]{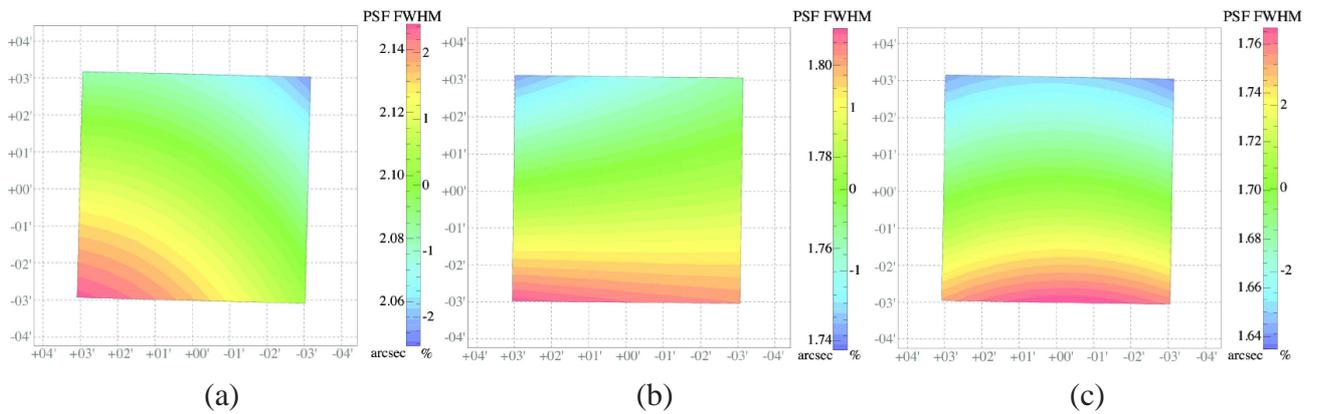} 
    \\
\hspace{-5mm} (a) \hspace{5cm} (b) \hspace{5cm} (c)
   \end{tabular}
   \end{center}
   \caption 
   { \label{fwhmm67} %>> use \label inside caption to get Fig. number with \ref{}
Same as Fig. \ref{psfm67}, but FWHM maps on the detectors.
} 
   \end{figure} 
%-------------

%------------- 
   \begin{figure}
   \begin{center}
   \begin{tabular}{c}
\hspace{-5mm}   \includegraphics[width=17cm]{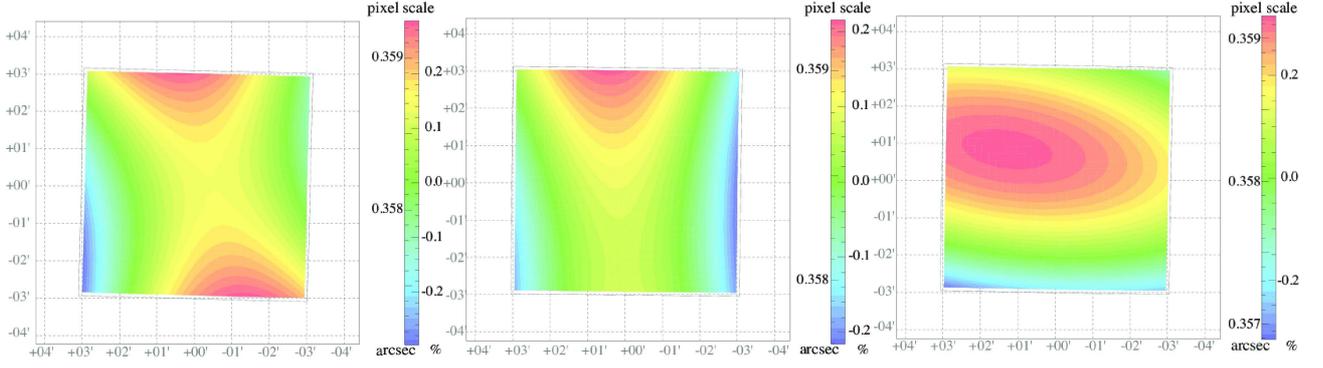} 
   \end{tabular}
   \end{center}
   \caption 
   { \label{distm67} %>> use \label inside caption to get Fig. number with \ref{}
Same as Fig. \ref{psfm67}, but pixel scale distortion maps on the detectors.
} 
   \end{figure} 
%-------------

We obtained images of the open cluster M67 with MuSCAT during the first light observation.
On this night, the sky condition was photometric and there was no moon.
The airmass of M67 was $\sim$1.2.
Using the M67 images, we have derived point spread function (PSF) and
distortion of MuSCAT images using IRAF, PSFEx\cite{2011ASPC..442..435B},
SExtractor\cite{1996A&AS..117..393B},
and SCAMP\cite{2006ASPC..351..112B}.
The extracted PSF, its ellipticity and FWHM of MuSCAT CCDs are shown in Figs. \ref{psfm67}, \ref{ellipm67},
and \ref{fwhmm67}.
We note that we have used "PIXEL$\_$AUTO" for the parameter of the PSF model in PSFEx.
The software estimates representative PSF for each grid using stars inside the grid.
We also note that the seeing in $g'_2$, $r'_2$, and $z_{s,2}$ was 2.1$''$, 1.8$''$, and 1.7$''$, respectively,
which was slightly worse than the typical seeing at the site of $\sim$1.5$''$ in optical bands.
Thus imaging quality was not limited by MuSCAT itself but by the seeing.
We have confirmed that the PSF is nearly circular throughout the FOV and that
MuSCAT does not have unexpected large aberration or imaging problems.

We have also derived distortion maps of images as differences in the pixel scale
on the CCDs using one of output options of SCAMP\cite{2004ASPC..314..551C}.
The derived pixel scale distortion maps are presented in Fig. \ref{distm67}.
The figure indicates that the pixel scale distortion is limited within about 0.3\%,
which is negligible for standard aperture photometry.

\subsection{Sensitivity and Efficiency}

\begin{figure}
   \begin{center}
   \begin{tabular}{c}
   \includegraphics[width=8.5cm]{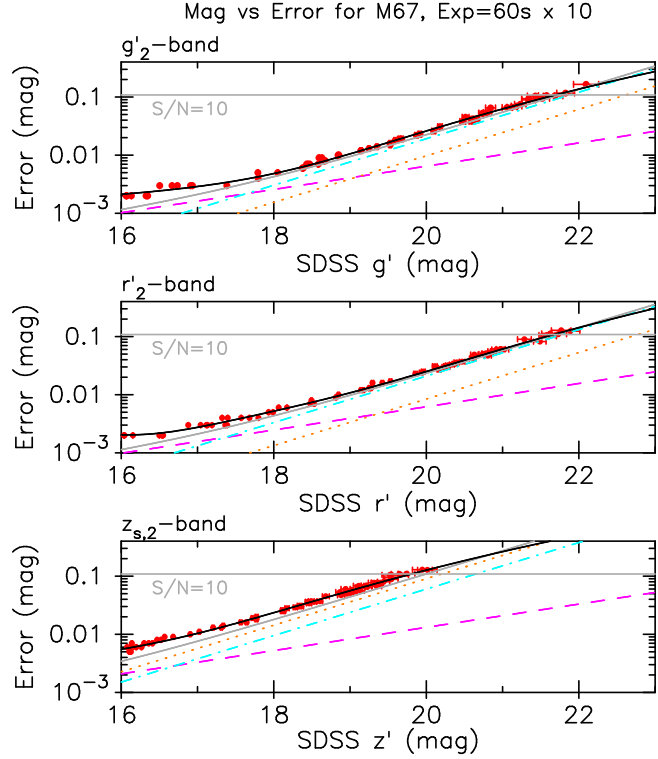} 
   \end{tabular}
   \end{center}
   \caption 
   { \label{limitmag} %>> use \label inside caption to get Fig. number with \ref{}
Relations of SDSS magnitudes and photometric errors for M67 stars taken with MuSCAT.
A dashed line in magenta, a dotted line in orange, a dashed-dotted line in cyan,
and a solid line in gray represent stellar photon noise,
readout noise, sky photon noise, and total (theoretical) noise, respectively.
The solid black line is the best-fit fifth-order function, just for visibility.
} 
   \end{figure}

We also estimate limiting magnitudes for the $g'_2$, $r'_2$, and $z_{s,2}$ bands using the images of M67.
The measured sky brightness in $g'_2$, $r'_2$, and $z_{s,2}$ were
19.9 mag arcsec$^{-2}$, 19.5 mag arcsec$^{-2}$, and 18.7 mag arcsec$^{-2}$, respectively.
For each band, 10 $\times$ 60 s images were obtained with dithering.
We note that we used high speed readout mode (2 MHz).
We apply bias-flat correction and stellar position alignment to the data and stacked them into a single image
for each band.
We conduct photometry for $\sim$100 stars on each stacked image
by using the DoPHOT package\cite{1993PASP..105.1342S},  which performs an analytical PSF fitting.
The measured instrumental magnitudes of these stars were then compared to
the SDSS 9 catalog\cite{2012ApJS..203...21A} for photometric calibration.
We note that we here ignore color terms and simply approximate that
the $g'_2$, $r'_2$, and $z_{s,2}$ bands are identical to the $g'$, $r'$, and $z'$ bands, respectively. 
Limiting magnitudes with 10-min exposure are estimated as the signal-to-noise (S/N) ratio reaches 10,
yielding $g'_\mathrm{lim} = 21.7$, $r'_\mathrm{lim} = 21.7$, and $z_\mathrm{lim} = 19.8$.
We note that $z_\mathrm{lim}$ is affected by higher readout noises in $z_{s,2}$ band
(see Table \ref{CCDsummary}) as we used the high speed readout mode.
For the S/N calculation, we simply adopt the photometric errors returned by DoPHOT.
We show a plot for SDSS magnitudes v.s. photometric errors in Fig. \ref{limitmag}. 

In addition, total throughput including the airmass, 188cm telescope, and MuSCAT
is measured by the same data.
We estimate total throughput as follows.
First, we measure the zero-point magnitudes on the stacked M67 images as
ZP($g'_2$)=28.63, ZP($r'_2$)=28.71, and ZP($z_{s,2}$)=27.08,
which correspond to 10 electrons for all bands.
Next, we estimate the expected incident flux coming from an astronomical object
with the above magnitudes
into the effective area of the primary mirror of the 188cm telescope.
Finally, comparing the expected flux with the detected one (10 electrons),
we estimate the total throughput
in $g'_2$, $r'_2$, and $z_{s,2}$ bands
as 20\%, 28\%, and 13\%, respectively.
The actual measured values are almost the same with expected values from the airmass,
188cm telescope, and MuSCAT, as shown in Table \ref{throughput}.

%------------- 
%-------------

%% Table 4

\subsection{High Precision Photometry}
\label{subsec:precision}

\begin{figure}
   \begin{center}
   \begin{tabular}{c}
   \includegraphics[width=8.5cm]{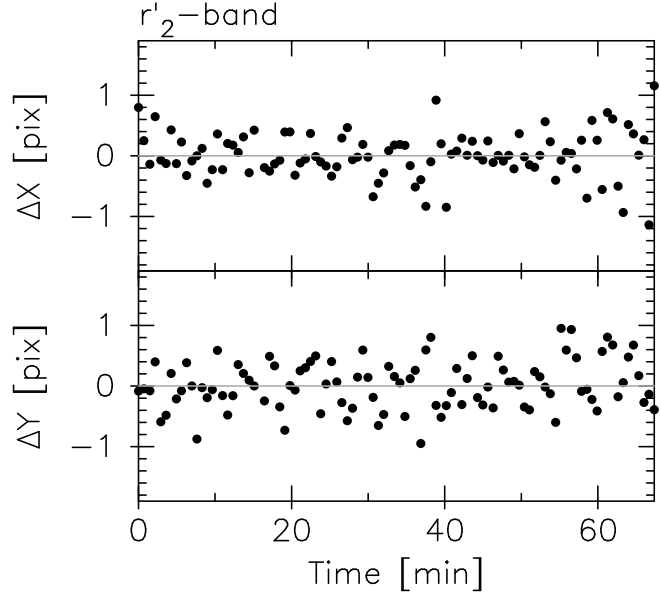} 
    \\
   \end{tabular}
   \end{center}
   \vspace{0.5cm}
   \caption 
   { \label{dxdy} %>> use \label inside caption to get Fig. number with \ref{}
Time series centroid positions of bright stars in the field of view for GJ~436.} 
   \end{figure}

In order to check for the photometric performance of MuSCAT,
we observed two stars; GJ 436\cite{2004ApJ...617..580B,2007A&A...472L..13G} and
WASP-12\cite{2009ApJ...693.1920H}, both hosting a transiting planet.
Observations were carried out on 2015 March 2 UT, during out-of-transit phases for both targets.
There was no cloud, but the sky level was relatively high due to a bright waxing moon with
the age (lunar phase) of 11.5.
We used the low speed readout mode.
GJ~436 is a nearby M3.5 dwarf with $g'$=11.81, $r'$=10.08, and $z'$=8.66 in the SDSS magnitudes.
GJ~436 was observed during 14:50--15:57 UT (airmass=1.05--1.01) with
the exposure time of 30 s for all bands.
WASP-12 is a G0 dwarf with $B$=12.1, $R$=11.2 (from the NOMAD 1.0 catalog\cite{2004AAS...205.4815Z}),
and $z'$=11.41\cite{2013MNRAS.428..182B}.
WASP-12 was observed during 13:37--14:39 UT (airmass=1.21--1.46) with
the exposure time of 30 s ($g'_2$ and $r'_2$) or 60 s ($z_{s,2}$). 
The exposure time and the number of exposures for each target, each filter is summarized in Table \ref{tbl:testobs}.
For each observation, the FOV was adjusted so that several similar-brightness stars
were simultaneously imaged.
The stellar image was defocused such that the FWHM of PSF became
$\sim$33--38 pixels and $\sim$31--36 pixels for GJ436 and WASP-12, respectively.
The self autoguiding system (see Sec. \ref{subsec:self}) was activated by
using $g'_2$-band images for both targets.
The stellar centroid changes in $r'_2$ band during the observations of GJ~436
are displayed in Fig. \ref{dxdy},
showing that the stellar positions were quite stable with the dispersion not exceeding $\sim$1 pixel.

\begin{figure}
   \begin{center}
   \begin{tabular}{c}
   \includegraphics[width=8.5cm]{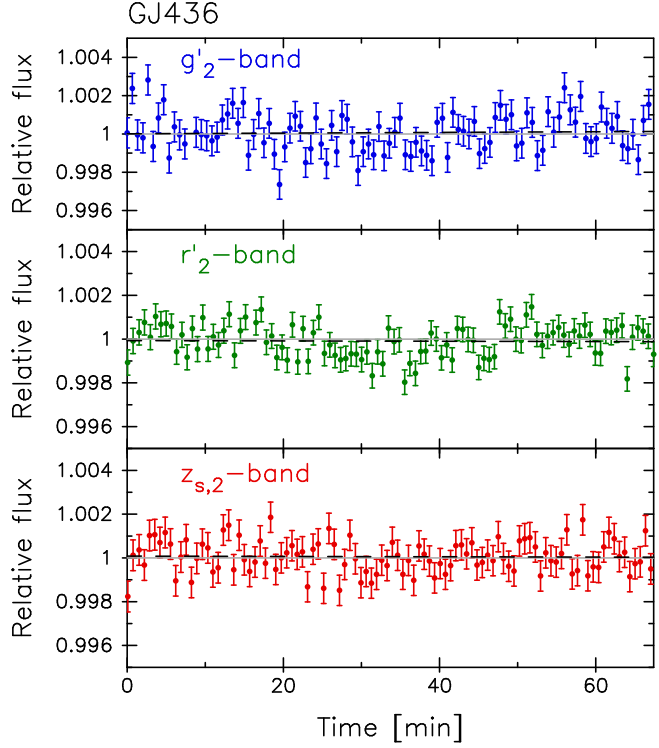} 
   \end{tabular}
   \end{center}
   \vspace{1cm}
   \caption 
   { \label{lcgj436} %>> use \label inside caption to get Fig. number with \ref{}
Light curves of GJ~436.} 
   \end{figure}

\begin{figure}
   \begin{center}
   \begin{tabular}{c}
   \includegraphics[width=8.5cm]{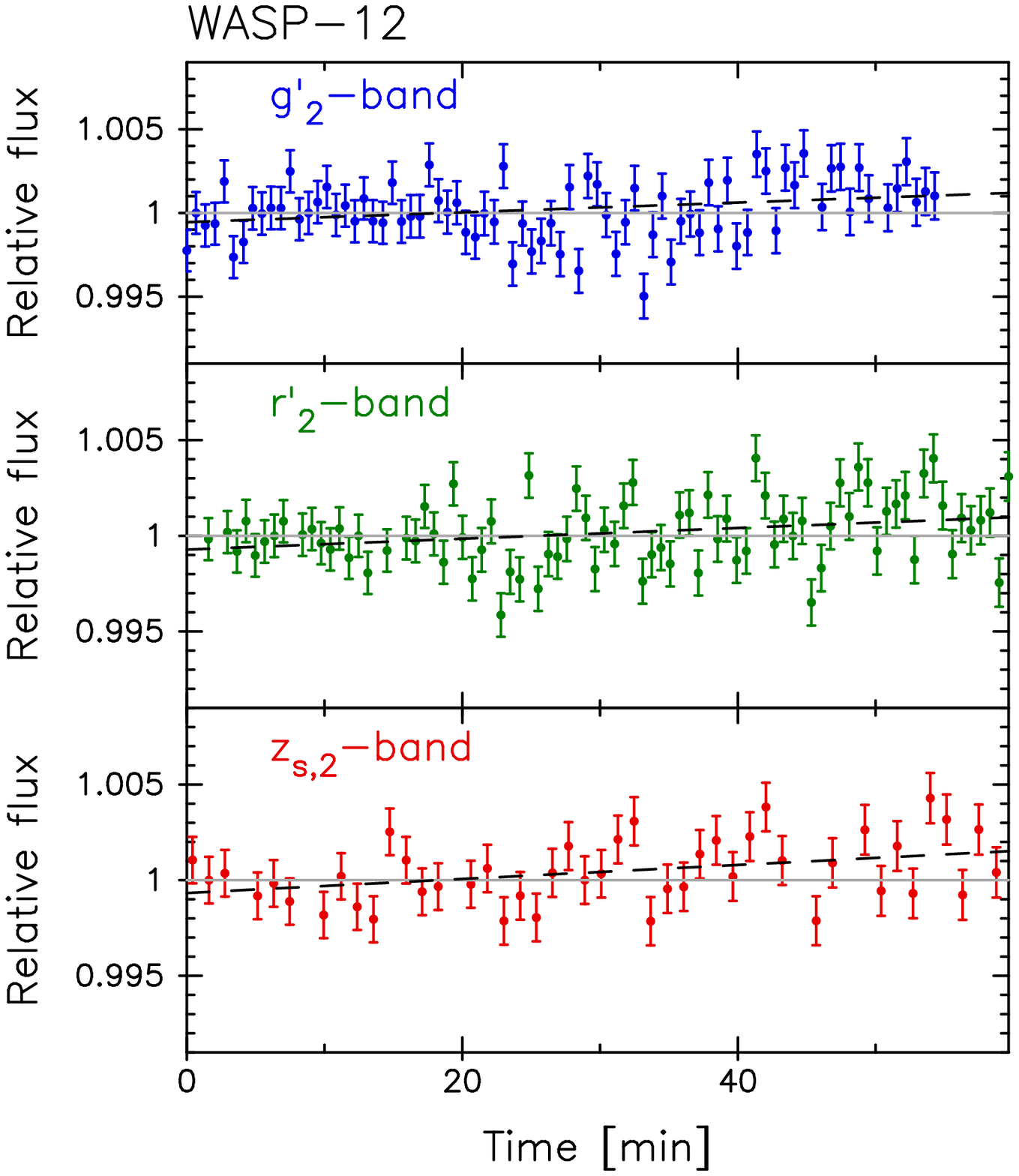} 
   \end{tabular}
   \end{center}
   \vspace{1cm}
   \caption 
   { \label{lcwasp12} %>> use \label inside caption to get Fig. number with \ref{}
Same as Fig. \ref{lcgj436}, but for WASP-12.} 
   \end{figure}

The observed data were reduced by using a customized aperture-photometry pipeline \cite{2011PASJ...63..287F}.
The applied aperture radius, the number of comparison stars used for the relative photometry,
and the unnormalized flux ratio of the target star and the ensemble of the comparison stars are summarized
in Table \ref{tbl:testobs}.
Note that the applied aperture radius was determined such that
the dispersion of the resultant light curve was minimum.
We show the resultant normalized light curves of GJ436 and WASP-12 in
Figs. \ref{lcgj436} and \ref{lcwasp12}, respectively.
The black dashed line indicates the best-fit linear function.
Photometric precisions, which we define as the root-mean-square (rms)
of the residual light curve from the liner fit,
achieve 0.101\%, 0.074\%, and 0.076\% in the $g'_2$, $r'_2$, and $z_{s,2}$ bands for GJ~436,
while those for WASP-12 are 0.16\%, 0.16\%, and 0.15\%, respectively.
These rms values are listed in Table \ref{tbl:testobs}.

%% Table 5

To see how the photometric performance of MuSCAT has been achieved,
we calculate the error budgets for these observations as shown in Table \ref{tbl:error_budget}.
In the table, $\sigma_\mathrm{target}$, $\sigma_\mathrm{comp}$, and $\sigma_\mathrm{sky}$
indicate photon noises arising from the target-star flux, comparison-star flux, and sky-background flux, respectively,
calculated assuming the Poisson (photon) noise.
$\sigma_\mathrm{read}$ is the noise contributed from the read-out noise listed in Table \ref{CCDsummary}.
$\sigma_\mathrm{scin}$ is the scintillation noise, for which we apply the following equation,
\begin{eqnarray}
\sigma_\mathrm{scin} = 0.064 D^{-2/3} (\sec Z)^{7/4} e^{-h/h_0} T^{-1/2},
\end{eqnarray}
where $D$ is the diameter of the primary mirror of the telescope in cm,
$Z$ is the zenith distance, $h=372$~m is the height above sea level of the observatory,
$h_0$ = 8000~m, and $T$ is the exposure time in seconds\cite{1967AJ.....72..747Y,1998PASP..110..610D}.
All the remaining (unknown, or difficult to assess) components of the photometric error are
treated as $\sigma_\mathrm{unknown}$, which is calculated as 
\begin{eqnarray}
\sigma_\mathrm{unknown} = \sqrt{\mathrm{rms}^2 - \sigma_\mathrm{target}^2 - \sigma_\mathrm{comp}^2 - \sigma_\mathrm{sky}^2 - \sigma_\mathrm{read}^2 - \sigma_\mathrm{scin}^2},
\end{eqnarray} where rms is the same as that listed in Table \ref{tbl:testobs}. 
Among these noise sources, all but $\sigma_\mathrm{unknown}$ are basically unavoidable.
Possible causes of $\sigma_\mathrm{unknown}$ can be the difference of
atmospheric transparency between toward the target star and toward the comparison stars,
the modulation of scintillation noise, the incompleteness of flat-field correction, and so on.
We indeed find that $\sigma_\mathrm{unknown}$ is the major noise source in some of the light curves,
but is still limited in degree of about 30--40\% in rms$^2$,
meaning that $\sigma_\mathrm{unknown}$ is not a very limiting factor for the photometric precision.
In other words, the most part of the photometric precision ($\gtrsim$ 60\% in rms$^2$ 
for all three bands) can be explained by the theoretical noise models.
We therefore consider that the expected photometric performance of MuSCAT has been well achieved.

We also note about the time-correlated noise (so-called the "red" noise) in the observed data.
For high-precision photometry such as transit observations,
a treatment of the red noise would be very important\cite{2006MNRAS.373..231P}.
We calculate a red-noise factor, which is the ratio of measured rms in binned data
to the one expected from the rms in unbinned data, for our observations.
We find ~1.3 in average, which is a typical value for ground-based transit observations.
Although one hour observations are not sufficient to evaluate the red noise in detail, we consider that
the level of the red noise of MuSCAT is similar to other ground-based instruments.
We will thus take into account the red noise for future science observations with MuSCAT.

%% Table 6

%------------- 
%------------- 

%------------- 
%------------- 

%------------- 
%------------- 

\section{Upgrading and Transferring Capability}
\label{sect:discussions}  % \label{} allows reference to this section

Although the current MuSCAT is ready for operations, it still has upgrading capabilities. 
First, MuSCAT can be upgraded in terms of FOV by replacing the three 1k$\times$1k CCD cameras,
which give 6.1$\times$6.1 arcmin$^2$ FOV,
with 2k$\times$2k CCD cameras, which will provide 12.7$\times$12.7 arcmin$^2$ FOV.
Such a wider FOV would be desirable to find good comparison stars
especially for very bright targets which will be discovered by the TESS mission.
Second, MuSCAT has space for another dichroic mirror to add NIR channels
(see Sec. \ref{subsec:overview}).
Additional NIR channels enable us to take images from optical to NIR simultaneously like GROND.
Such a capability will enhance scientific merits for transmission spectroscopy in the light of
efficiency and simultaneity. 

MuSCAT also has a transferring capability.
The current instrument is optimized for the 188cm telescope at Okayama Astrophysical Observatory
whose F-number is F18, but MuSCAT can be transferred to or can make a copy of itself for
other telescopes by replacing F conversion lenses.

\section{Summary}
\label{sect:summary}  % \label{} allows reference to this section

We have developed a new astronomical instrument MuSCAT for the 188cm telescope
at Okayama Astrophysical Observatory in Japan.
MuSCAT has a capability of 3-color simultaneous imaging in 
$g'_2$ (400--550 nm), $r'_2$ (550--700 nm), and $z_{s,2}$ (820--920 nm) bands
with three 1k$\times$1k pixel CCDs.
The field of view of MuSCAT is 6.1$\times$6.1 arcmin$^2$
with the pixel scale of 0.358 arcsec per pixel.

One of the prime aims of MuSCAT is to confirm whether candidates of transiting planets
discovered by transit surveys, including such as K2, TESS, and PLATO,
are truly planets or false positives due to eclipsing binaries.
Another prime aim of MuSCAT is to measure the wavelength dependence of transit depths in visible bands,
providing rough information about exoplanetary atmospheres:
such as the feature of the Rayleigh scattering by hydrogen dominated atmospheres,
the feature of the Mie scattering by hazy atmospheres, or the flat feature of cloudy
atmospheres\cite{2012ApJ...756..176H,2013ApJ...770...95F}.

The capability of multi-color simultaneous transit photometry is well suitable for those aims.
Since MuSCAT can achieve 0.1\% photometric precision with 30 s exposure
for stars brighter than $\sim$10 mag
as shown in Sec. \ref{subsec:precision}, MuSCAT will work effectively for the purposes
especially for bright TESS transiting planets.
In addition, MuSCAT would be also useful for follow-up observations of supernovae and
gamma-ray bursts, and monitoring variable stars, and so on.
The instrument is ready for operation at Okayama Astrophysical Observatory.

\bigskip
\acknowledgments 

We are very grateful for kind helps by the staffs of Okayama Astrophysical Observatory,
Astrobiology Center, and Exoplanet Detection Project Office in NAOJ.
We acknowledge the following people who helped us to obtain a research grant to
develop this instrument:
Masahiro Ikoma, Yui Kawashima, Bun'ei Sato, Yasuhito Sekine,
Hidenori Genda, Takahiro Nagayama, Keigo Enya, and Mikio Kurita.
N. N. acknowledges supports by the NAOJ Fellowship,
Inoue Science Research Award, NINS Program for Cross-Disciplinary Study,
and Grant-in-Aid for Scientific Research (A) (JSPS KAKENHI Grant Number 25247026).
A.F. is supported by the Astrobiology Project of the Center for Novel Science Initiatives (CNSI),
National Institutes of Natural Sciences (NINS) (Grant Number AB261005).
N.K. is supported by NINS Program for Cross-Disciplinary Study.
M.T. was partly supported by JSPS KAKENHI Grant Number 22000005.

\clearpage
%%%%%%%%%%%%%%%%%%%%%%%%%%%%%%%%%%%%%%%%%%%%%%%%%%%%%%%%%%%%%
%%%%% References %%%%%

%\bibliography{refs}   %>>>> bibliography data in report.bib
\bibliographystyle{spiejour}   %>>>> makes bibtex use spiejour.bst

%%%%%%%%%%%%%%%%%%%%%%%%%%%%%%%%%%%%%%%%%%%%%%%%%%%%%%%%%%%%%
%%%%% Biographies of authors %%%%%

\vspace{2ex}\noindent{\bf Norio Narita} is the principal investigator of the MuSCAT instrument and its observing team.
He is jointly appointed from Astrobiology Center, National Astronomical Observatory of Japan,
and SOKENDAI (The Graduate University of Advanced Studies) as a research assistant professor.

\vspace{1ex}
\noindent Biographies and photographs of the other authors are not available.

%%%%%%%%%%%%
\begin{table}
\caption{
\label{CCDsummary}
Summary of basic information of cameras and CCDs
}
\begin{center}
\begin{tabular}{clll}
\hline \hline
\multicolumn{2}{c}{Camera name}  & PIXIS 1024B / PIXIS 1024B\_eXcelon \\
\hline

\multicolumn{2}{c}{CCD format} & \multicolumn{2}{l}{1024 $\times$ 1024 pixels}\\

\multicolumn{2}{c}{Pixel size} & \multicolumn{2}{l}{13 $\times$ 13 $\mu$m}\\

\multicolumn{2}{c}{Imaging area} & \multicolumn{2}{l}{13.3 $\times$ 13.3 mm}\\

\multicolumn{2}{c}{Data interface} & \multicolumn{2}{l}{USB2.0}\\

\multicolumn{2}{c}{Dimensions} & \multicolumn{2}{l}{16.59 cm $\times$ 11.81 cm $\times$ 11.38 cm (L $\times$ W $\times$ H) }\\ 

\multicolumn{2}{c}{Weight} & \multicolumn{2}{l}{2.27 kg }\\ 

\hline

\multicolumn{2}{c}{Selectable gain modes}& \multicolumn{2}{l}{1, 2, 4 e$^-$/ADU}\\

 & mode 1 & \multicolumn{2}{l}{1.00 (ch 1), 1.02 (ch 2), 1.00 (ch 3) e$^-$/ADU}\\
Measured gains & mode 2 & \multicolumn{2}{l}{1.98 (ch 1), 2.02 (ch 2), 2.00 (ch 3) e$^-$/ADU}\\
 & mode 4 & \multicolumn{2}{l}{3.96 (ch 1), 4.05 (ch 2), 3.97 (ch 3) e$^-$/ADU}\\

\multicolumn{2}{c}{Dark current @ $-70^\circ$C}& \multicolumn{2}{l}{$<$0.0015 e$^-$s$^{-1}$pix$^{-1}$ (measured)}\\
&& \multicolumn{2}{l}{0.0004 e$^-$s$^{-1}$pix$^{-1}$ (catalog typical value)}\\

Full well: & Single pixel & \multicolumn{2}{l}{$\sim$70000 e$^-$}\\

\multicolumn{2}{c}{Selectable readout speed (readout time)}& \multicolumn{2}{l}{100 kHz (10.0 s), 2 MHz (0.58 s)}\\

Measured read noise: & @100 kHz & \multicolumn{2}{l}{3.9 (ch 1), 3.8 (ch 2), 4.2 (ch 3) e$^-$}\\

 & @2 MHz & \multicolumn{2}{l}{11 (ch 1), 12 (ch 2), 24-27$^\mathrm{a}$ (ch 3) e$^-$}\\

\hline
\end{tabular}
\\
{\scriptsize
$^\mathrm{a}$: An additional pattern (not random) noise is seen only in the 2MHz mode of the ch 3 CCD.
This is scheduled to be repaired (the repair has been completed at the time of publication).
}
\end{center}
\end{table}
%%%%%%%%%%%%

%%%%%%%%%%%%
\begin{table}
\caption{
\label{totalefficiencytable}
Wavelength dependence of expected total efficiencies of MuSCAT.
}
\begin{center}
\begin{tabular}{cccccccc}
\hline
wavelength(nm) & $g'_2$(\%) & $r'_2$(\%)  & $z_{s,2}$(\%)  & wavelength(nm) & $g'_2$(\%) & $r'_2$(\%)  & $z_{s,2}$(\%) \\
\hline
360	&	0.00 	&	0.00 	&	0.00 	&	660	&	0.00 	&	55.37 	&	0.00 	\\
370	&	0.00 	&	0.00 	&	0.00 	&	670	&	0.00 	&	54.80 	&	0.00 	\\
380	&	0.00 	&	0.00 	&	0.00 	&	680	&	0.00 	&	54.35 	&	0.00 	\\
390	&	0.00 	&	0.00 	&	0.00 	&	690	&	0.00 	&	52.67 	&	0.00 	\\
400	&	2.02 	&	0.00 	&	0.00 	&	700	&	0.00 	&	0.07 	&	0.00 	\\
410	&	42.21 	&	0.00 	&	0.00 	&	710	&	0.00 	&	0.00 	&	0.00 	\\
420	&	44.63 	&	0.00 	&	0.00 	&	720	&	0.00 	&	0.00 	&	0.00 	\\
430	&	46.75 	&	0.00 	&	0.00 	&	730	&	0.00 	&	0.00 	&	0.00 	\\
440	&	48.39 	&	0.00 	&	0.00 	&	740	&	0.00 	&	0.00 	&	0.00 	\\
450	&	50.36 	&	0.00 	&	0.00 	&	750	&	0.00 	&	0.00 	&	0.00 	\\
460	&	50.99 	&	0.00 	&	0.00 	&	760	&	0.00 	&	0.00 	&	0.00 	\\
470	&	52.42 	&	0.00 	&	0.00 	&	770	&	0.00 	&	0.00 	&	0.00 	\\
480	&	52.35 	&	0.00 	&	0.00 	&	780	&	0.00 	&	0.00 	&	0.00 	\\
490	&	53.23 	&	0.00 	&	0.00 	&	790	&	0.00 	&	0.00 	&	0.00 	\\
500	&	53.56 	&	0.00 	&	0.00 	&	800	&	0.00 	&	0.00 	&	0.00 	\\
510	&	54.21 	&	0.00 	&	0.00 	&	810	&	0.00 	&	0.00 	&	0.00 	\\
520	&	54.45 	&	0.00 	&	0.00 	&	820	&	0.00 	&	0.00 	&	0.08 	\\
530	&	54.87 	&	0.00 	&	0.00 	&	830	&	0.00 	&	0.00 	&	35.05 	\\
540	&	54.58 	&	0.00 	&	0.00 	&	840	&	0.00 	&	0.00 	&	36.12 	\\
550	&	15.81 	&	0.00 	&	0.00 	&	850	&	0.00 	&	0.00 	&	34.50 	\\
560	&	0.00 	&	0.06 	&	0.00 	&	860	&	0.00 	&	0.00 	&	32.74 	\\
570	&	0.00 	&	57.17 	&	0.00 	&	870	&	0.00 	&	0.00 	&	30.78 	\\
580	&	0.00 	&	56.60 	&	0.00 	&	880	&	0.00 	&	0.00 	&	29.08 	\\
590	&	0.00 	&	57.07 	&	0.00 	&	890	&	0.00 	&	0.00 	&	27.13 	\\
600	&	0.00 	&	56.74 	&	0.00 	&	900	&	0.00 	&	0.00 	&	25.21 	\\
610	&	0.00 	&	56.73 	&	0.00 	&	910	&	0.00 	&	0.00 	&	22.99 	\\
620	&	0.00 	&	56.56 	&	0.00 	&	920	&	0.00 	&	0.00 	&	14.40 	\\
630	&	0.00 	&	56.29 	&	0.00 	&	930	&	0.00 	&	0.00 	&	0.02 	\\
640	&	0.00 	&	55.97 	&	0.00 	&	940	&	0.00 	&	0.00 	&	0.00 	\\
650	&	0.00 	&	55.63 	&	0.00 	&	950	&	0.00 	&	0.00 	&	0.00 	\\
\hline
\end{tabular}
\end{center}
\end{table}
%%%%%%%%%%%%

%------------- 

\begin{table}
\caption{
\label{linearity}
Summary of linearity information of CCDs for each gain and each readout speed.
}
\begin{center}
\begin{tabular}{ccccc}
\hline \hline
Gain mode&ADC speed [Hz]& $g'_2$-band & $r'_2$-band & $z_{s,2}$-band\\
&&linearity slope [\%]&inearity slope [\%]&linearity slope [\%]\\
\hline
1&$2\times 10^6$& $0.154 \pm 0.062$ & $0.104 \pm 0.051$ & $0.204 \pm 0.026$ \\
&$1\times 10^5$& $-0.198 \pm 0.071$ & $-0.055 \pm 0.043$ & $-0.133 \pm 0.040$ \\
2&$2\times 10^6$& $0.016 \pm 0.072$ & $0.018 \pm 0.045$ & $0.112 \pm 0.025$ \\
&$1\times 10^5$& $-0.210 \pm 0.069$ & $0.005 \pm 0.043$ & $-0.061 \pm 0.039$ \\
4&$2\times 10^6$& $0.169 \pm 0.044$ & $0.120 \pm 0.037$ & $0.115 \pm 0.030$ \\
&$1\times 10^5$& $-0.059 \pm 0.077$ & $-0.085 \pm 0.043$ & $-0.037 \pm 0.034$ \\
\hline
\end{tabular}
\end{center}
\end{table}
%------------- 

%-------------
\begin{table}
\begin{center}
{\footnotesize
\caption{Summary of throughput (TP) of MuSCAT on the 188cm telescope at OAO}
\label{throughput}
\scalebox{0.9}{
\begin{tabular}{ccccccc}\hline
band& atmos. transmittance & M1 & M2 & MuSCAT & expected total TP & measured TP\\ \hline
$g'_2$ &$\sim$60\%$^a$ & 86\%$^b$ & 79\%$^b$ & 51\%$^d$ & 21\% &20\% \\
$r'_2$ &$\sim$65\%$^a$ & 85\%$^b$ & 85\%$^b$ & 56\%$^d$ & 27\% &28\% \\
$z'_{s,2}$ &$\sim$80\%$^a$ & $\sim$80\%$^c$ & $\sim$80\%$^c$ & 29\%$^d$ & 15\% & 13\% \\ \hline
\end{tabular}
}}
{\scriptsize \\ M1 = main mirror, M2 = secondary mirror, MuSCAT = all optics * BBAR coating * QE,\\
$^a$: typical, $^b$: measured, $^c$: extrapolated from measured data up to 750 nm,
$^d$: expected}
\end{center}
\end{table}
%-------------

%%
\begin{table}[t]
\begin{center}
\caption{Summary of the test observation and analysis
\label{tbl:testobs}}
\vspace{5pt}
\small
\begin{tabular}{lcccccc}
\hline
\hline
Target (Filter) & Exp. time [s] & $N_\mathrm{data}$ $^\mathrm{a}$ & $N_\mathrm{comp}$ $^\mathrm{b}$ & $F_\mathrm{t}/F_\mathrm{c}$ $^\mathrm{c}$ & $R_\mathrm{ap}$ $^\mathrm{d}$ [pixel]  & rms $^\mathrm{e}$ [\%]  \\
\hline
GJ436 ($g'_2$) & 30 & 99 & 2 & 0.42 & 24 & 0.101\\
GJ436 ($r'_2$) & 30 & 99 & 3 & 0.85 & 26 & 0.074\\
GJ436 ($z'_{s,2}$) & 30 & 100 & 2 & 3.6 & 24  & 0.076\\
WASP-12 ($g'_2$) & 30 & 79 & 4 & 0.38 & 22 & 0.16\\
WASP-12 ($r'_2$) & 30 & 82 & 3 & 0.52 & 24 & 0.16\\
WASP-12 ($z'_{s,2}$) & 60 & 45 & 2 & 0.59 & 22 & 0.15\\
\hline
\end{tabular}
\end{center}
{\scriptsize
$^\mathrm{a}$ The number of observed data points.\\
$^\mathrm{b}$ The number of comparison stars used for relative photometry.\\
$^\mathrm{c}$ Unnormalized flux ratio of the target star and the ensemble of the comparison stars.\\
$^\mathrm{d}$ Applied aperture radius.\\
$^\mathrm{e}$ The rms value of the residual light curve from a linear fit.\\
}
\end{table}
\begin{table}[t]
\begin{center}
\small
\caption{Error budget
\label{tbl:error_budget}}
\vspace{5pt}
\begin{tabular}{lccccccc}
\hline
\hline
Target (Filter) & $\sigma_\mathrm{target}$ [\%]& $\sigma_\mathrm{comp}$ [\%]& $\sigma_\mathrm{sky}$ [\%]& $\sigma_\mathrm{read}$ [\%]& $\sigma_\mathrm{scin}$ [\%]& $\sigma_\mathrm{unknown}$ [\%]& rms [\%]\\ 
\hline
WASP-12 ($g'_2$) & 0.058 & 0.036 & {\bf 0.092}$^\mathrm{a}$ & 0.005 & 0.078 & 0.076 & 0.16\\
WASP-12 ($r'_2$) & 0.047 & 0.034 & 0.077 & 0.004 & 0.081 & {\bf 0.106}$^\mathrm{a}$ & 0.16\\
WASP-12 ($z_{s,2}$) & 0.065 & 0.048 & {\bf 0.083}$^\mathrm{a}$ & 0.006 & 0.057 & {\bf 0.083}$^\mathrm{a}$ & 0.15\\
GJ436 ($g'_2$) & 0.043 & 0.028 & 0.037 & 0.003 & 0.050 & {\bf 0.061}$^\mathrm{a}$ & 0.101\\
GJ436 ($r'_2$) & 0.023 & 0.021 & 0.019 & 0.001 & {\bf 0.050}$^\mathrm{a}$ & 0.041 & 0.074\\
GJ436 ($z_{s,2}$) & 0.023 & 0.043 & 0.023 & 0.003 & {\bf 0.050}$^\mathrm{a}$ & 0.019 & 0.076\\
\hline
\end{tabular}
\end{center}
{\footnotesize
$^\mathrm{a}$ The bold text indicates the most dominant noise component for each light curve.
}
\end{table}
%%

%\listoftables

\end{spacing}
\end{document}